\documentclass[12pt]{article}
\usepackage{amsfonts}
\usepackage{amssymb}
\usepackage{graphics}
\usepackage{epsfig}
\usepackage{multirow}

\usepackage{latexsym}
\usepackage{amsmath}

\usepackage{relsize}
\usepackage{geometry}
\usepackage{color}

\newcommand{\beq}{\begin{equation}}
\newcommand{\eeq}{\end{equation}}
\newcommand{\beqn}{\begin{eqnarray}}
\newcommand{\eeqn}{\end{eqnarray}}
\newcommand{\bea}{\begin{eqnarray}}
\newcommand{\eea}{\end{eqnarray}}
\newcommand{\beas}{\begin{eqnarray*}}
\newcommand{\eeas}{\end{eqnarray*}}

\newcommand{\bquo}{\begin{quote}}
\newcommand{\enqu}{\end{quote}}

\def\Tr{ \hbox{\rm Tr}}

\def\im{\hbox{\rm Im}}

\def\Re{\hbox {\rm Re}}
\def\Im{\hbox {\rm Im}}

\newcommand{\gsim}{\lower.7ex\hbox{$\;\stackrel{\textstyle>}{\sim}\;$}}
\newcommand{\lsim}{\lower.7ex\hbox{$\;\stackrel{\textstyle<}{\sim}\;$}}

\def\stroke{\vrule height8pt width0.4pt depth-0.1pt}
\def\topfleck{\vrule height8pt width0.5pt depth-5.9pt}
\def\botfleck{\vrule height2pt width0.5pt depth0.1pt}
\def\Zmath{\vcenter{\hbox{\numbers\rlap{\rlap{Z}\kern 0.8pt\topfleck}\kern
2.2pt   \rlap Z\kern 6pt\botfleck\kern 1pt}}}
\def\Qmath{\vcenter{\hbox{\upright\rlap{\rlap{Q}\kern
                   3.8pt\stroke}\phantom{Q}}}}
\def\Nmath{\vcenter{\hbox{\upright\rlap{I}\kern 1.7pt N}}}
\def\Cmath{\vcenter{\hbox{\upright\rlap{\rlap{C}\kern
                   3.8pt\stroke}\phantom{C}}}}
\def\Rmath{\vcenter{\hbox{\upright\rlap{I}\kern 1.7pt R}}}
\def\Z{\ifmmode\Zmath\else$\Zmath$\fi}
\def\Q{\ifmmode\Qmath\else$\Qmath$\fi}
\def\N{\ifmmode\Nmath\else$\Nmath$\fi}
\def\C{\ifmmode\Cmath\else$\Cmath$\fi}
\def\R{\ifmmode\Rmath\else$\Rmath$\fi}

\def\Tr{{\rm Tr}}

\def\Im{{\rm Im}}
\def\Re{{\rm Re}}

\def\2{{1\over 2}}

\def\4N{${\cal N}=4$}
\def\N{{\mathcal N}}

\def\beq{\begin{equation}}
\def\eeq{\end{equation}}
\def\ba{\beq\new\begin{array}{c}}
\def\ea{\end{array}\eeq}

\def\Tr{ \hbox{\rm Tr}\,}

\def\R{{\rm R}}

\def\im{\hbox{\rm Im}}

\def\Re{\hbox {\rm Re}\,}
\def\Im{\hbox {\rm Im}\,}

\def\Z{\mathrm Z}
\def\1{\mathbbm{1}}
\def\N{{\cal N}}

\def\C{\rm C}

\def\1{\mbox{\tiny (1) }}
\def\0{\mbox{\tiny (0) }}

\begin{document}

\begin{titlepage}

\begin{flushright}
June 15, 2010
\end{flushright}

\vspace{1mm}

\begin{center}
{\large  {\bf On metastable vacua  \\[4mm] 
in perturbed \boldmath{$\mathcal{N}=2$} theories }}
\end{center}

\vspace{1.0mm}

\begin{center}
{\large   {\sc Roberto Auzzi}$^{(1)}$} and {\large  {\sc Eliezer Rabinovici}$^{(2)}$} 

\vspace{10.0mm}

{\it \footnotesize
Racah Institute of Physics, The Hebrew University,  \\ Jerusalem 91904, Israel}
\\[4mm]
$^{(1)}$ {\tt auzzi@phys.huji.ac.il} \\
$^{(2)}$ {\tt eliezer@vms.huji.ac.il} 
\end{center}

\vspace{10.0mm}

\begin{abstract}

We study supersymmetry breaking in metastable vacua 
on the Coulomb branch of  perturbed $\mathcal{N}=2$ gauge theories,
with gauge group $SU(2)$ and  different matter content ($N_f=0,2,4$).
The theory is deformed with a superpotential which is a
cubic  polynomial in $u=\Tr \Phi^2$,
where $\Phi$ is the adjoint superfield. 
 The allowed region of the perturbation parameters
 in this $\mathcal{N}=1$ theory
is plotted  as a function of the moduli space coordinate.
 In the asymptotically free cases a significant fine-tuning
 in the perturbation parameters is needed to achieve metastable vacua in the weakly coupled 
 region of the moduli space; a lower degree of
 fine-tuning is required in the strongly coupled regime.
In the conformal case  ($N_f=4$ fundamentals)
we find that also an explicit mass for the hypermultiplets
must be introduced in order to generate metastable vacua. 
In the case of  $N_f=2$ fundamentals 
 it is possible to achieve a metastable vacuum also in the
neighborhood of the Argyres-Douglas fixed point
(even if a large degree of fine-tuning is needed in this limit).  
Direct gauge mediation is discussed; gaugino masses 
of the same order of the SUSY-breaking can be obtained.

\end{abstract}

\end{titlepage}

\vfill\eject

\section{Introduction}

Long-lived metastable vacua which break supersymmetry
are  generic  in $\mathcal{N}=1$
 theories with massive fundamental matter,
 as was shown by ISS \cite{ISS}
(see \cite{ISreview,KOOreview} for
 reviews of the topic). Usually the strong coupling of the models
inhibits reliable calculations in
 models of dynamical symmetry breaking. 
In the ISS setting this is avoided
using the Seiberg dual description \cite{Sdual},
which is weakly coupled even in some cases where the original theory
is strongly coupled. This is true in particular  for $\mathcal{N}=1$ SQCD with $N_c<N_f<3/2 N_c$;
in this regime the dual description is that of a an $SU(N_f-N_c)$ gauge group 
with $N_f$ fundamentals and some scalars. This theory is infrared free
and computations are reliable, they lead to parametrically long lived meta stable states. 
The study of metastability inside the conformal window 
($3/2 N_c<N_f<3 N_c$) is more involved,
also because the dual theory is not weakly coupled\,\footnote{
Recently examples of  metastable vacua  in the conformal window
of $\mathcal{N}=1$  SQCD
 (where some of the flavors are coupled to 
a gauge singlet) have been discussed in \cite{ITYY} and  \cite{AGMS}.}.

Another setting in which dynamical supersymmetry
breaking  in a long-lived metastable vacua is calculable is in
 $\mathcal{N}=2$ theories, perturbed by a small superpotential.
 From the Seiberg-Witten curves \cite{sw1,sw2},
 the low energy effective theory on the Coulomb branch is exactly known, 
 including the K\"ahler potential. 
 Using this theoretical tool, in \cite{OOP,MOOP,pastras1,pastras2}
 the issue of metastability was studied in $\mathcal{N}=2$ theories
 perturbed by a superpotential\,\footnote{ Metastable vacua in $\mathcal{N}=2$
theories perturbed by a Fayet-Iliopoulos term were studied in \cite{amos}.}.
These vacua have been also realized in string/M-theory 
constructions in \cite{others}. 

Consider the case of the $\mathcal{N}=2$ theory with
gauge group $SU(2)$, with arbitrary
matter content consistent with asymptotic freedom;
the Coulomb branch of the moduli space can be parameterized 
in this case by the coordinate $u=\Tr \Phi^2$. 
The moduli space can be lifted by perturbing the theory with
a small superpotential which is a function of $u$: 
 \beq
\mathcal{W} = \mu \left( u + \alpha u^2 + \beta u^3   \right) \, . \label{tumiperturbi}
 \eeq
The term linear in $u$ is a mass term
for the adjoint superfield; the terms proportional 
to $u^2$ and $u^3$ correspond instead to  (dangerously) irrelevant operators.
In \cite{OOP} it was shown that for almost every point
of the moduli space $u_0$, it is possible to choose
the coefficients $(\alpha,\beta)$ in (\ref{tumiperturbi}) in such a way that 
a metastable vacuum is generated at $u=u_0$. 

The direct computation of the superpotential which is needed 
to make each point of the moduli space metastable 
 involves some rather cumbersome expressions 
for the $\mathcal{N}=2$ prepotential (this is true especially
in the cases with matter hypermultiplets).
In \cite{OOP} the allowed range of the deformation parameters
allowing a metastable vacuum  at the origin of the moduli space
was explicitly computed 
 in the pure Super-Yang-Mills case, for a generic number of colors.
Explicit expressions for the case of a generic point in the moduli space in $\mathcal{N}=2$
Super-Yang-Mills with gauge group $SU(2)$ were found in \cite{pastras2}. 
In this paper we  compute numerically the allowed region
for the deformation parameters $(\alpha,\beta)$
in some cases that were not discussed before.

In particular we focus on nearly scale invariant  theories. Theories which are scale  invariant do not have metastable
states. This is true whether the symmetry is spontaneously  broken or not.
Using scale invariance, any candidate for a metastable state can be scaled
to zero energy. In other words, no scale is available to produce the
local stability around the metastable state.
 Scale invariant and conformal theories have many interesting properties not the least of them
is the control on the value of vacuum energy \cite{scales}.

The examples that we discuss for nearly scale invariant theories are 
 the  $\mathcal{N}=2$ $SU(2)$  theory with $N_f=4$ fundamental hypermultiplets
and the $\mathcal{N}=4$ theory. When the hypermultiplets are massless, 
it is not possible  to generate a metastable vacuum at any point of the moduli space 
 with the perturbation (\ref{tumiperturbi}).
The situation changes once one  introduces a mass $m$
for some of the hypermultiplets; then the result in \cite{OOP}
applies and it is possible to generate metastable vacua.

Another case that we  study is the one with $N_f=2$
fundamental massive hypermultiplets. For a critical value of the
hypermultiplet mass, an Argyres-Douglas \cite{adapsw} conformal vacuum appears 
in the moduli space.
It turns out that generating a metastable vacuum in the neighborhood of
the conformal point is especially difficult,
 $(\alpha,\beta)$ must be rather fine-tuned.
In this specific example we find that
the allowed parameter range
vanish as $(u_0-u_{AD})^3$, which is much stronger than
nearby other supersymmetric vacua, where we find
it vanishes in vanishes as $(u_0-u_{susy})$.

The examples that we consider in this paper contain
flavor symmetries that can be gauged and coupled
to external supersymmetric sectors, 
in order to realize direct gauge mediation
 (see for example \cite{ogm-review,KOOreview} for reviews).
 The gaugino masses obtained are of the same order of the SUSY-breaking.
  In particular, if we consider theories with zero mass term for the hypermultiplets,
 ordinary gauge mediation  is realized.

The paper is organized as follows. In section 2 we review
the theoretical setting and
 the general strategy to compute the range of parameters  of $(\alpha,\beta)$
 in order to generate a metastable vacuum.  This gives a sense of the genericity  of forming metastable states.
 In section 3 we discuss the $\mathcal{N}=2$ Super-Yang-Mills theory ($N_f=0$). 
 In section 4 the conformal cases (the theory with $N_f=4$ fundamentals
 and $\mathcal{N}=4$ SYM) are studied.
 Section 5 is about the theory with $N_f=2$  fundamentals which has conformal points.
In all cases we search and find parametrically long lived metastable states.
 In section 6 we comment about direct gauge mediation.
Section 7 contains the conclusions. 
The appendix  concerns the weakly coupled limit, 
where a compact analytical expression
for the  range of parameters of  $(\alpha,\beta)$
 can be found.

\section{Theoretical setting}

Consider an $\mathcal{N}=2$ theory
with gauge group $SU(2)$ and arbitrary matter
content consistent with asymptotic freedom or conformal invariance.
The moduli space can be parameterized by the VEV
\beq u = \Tr \Phi^2 \, ,\eeq
which spontaneously breaks the $SU(2)$ gauge symmetry to $U(1)$.
The low-energy dynamics is described by the Seiberg-Witten curve  \cite{sw1,sw2}, 
which enables to compute the K\"ahler potential of the low energy
effective $U(1)$ theory.
The result is expressed in term of the functions $a(u),a_D(u)$;
 $\tau_e$ is defined as $\tau_e=\frac{d a_D}{d a}$.
The following convention is used for
 the effective $U(1)$ coupling $g_e$ and $\theta$ angle:
\beq \tau_e=\frac{\theta}{\pi} + \frac{8 \pi \, i }{g_{e}^2} \, . \eeq
The K\"ahler metric on the moduli space is
given in term of the holomorphic functions $a(u),a_D(u)$:
\beq ds^2=(\im \tau_e) \, d a \, d \bar{a} \, =
g \, du \, d \bar{u} \, , \qquad
 g= (\im \, \tau_e) a' \, \bar{a}' =\Im \left( a_D' \, \bar{a}' \right) 
\, ,\eeq
where $a'=da/du$ and $\bar{a}'=d \bar{a}/ d \bar{u}$.
Spontaneous supersymmetry breaking 
in  metastable vacua is generated by deforming
the theory with a superpotential $\mathcal{W}(u)$.
The potential on the moduli space is then
\beq V=\frac{|\mathcal{W}'(u)|^2}{g} \, .\eeq

In \cite{OOP} it is shown that by an appropriate choice
for the superpotential it is indeed possible to generate a
metastable vacuum in almost every point of the moduli space;
the proof relies on the fact that any sectional curvature
of the Riemann curvature tensor $R$ of the moduli space metric
 is strictly positive definite in almost
every point of the moduli space.
This means that (with the exception of a finite number
of points in the moduli space)
for any two vectors $w_1,w_2$ on the tangent space,
\[ \langle w_1,R(w_2,w_2) w_1 \rangle > 0 \, ,\]
for every $w_1,w_2 \neq 0$.

The task of finding metastable vacua is equivalent
to finding a local maximum of 
\beq \frac{1}{V} = \Im (\tau_e)  \left| \frac{a'}{\mathcal{W}'} \right|^2 \, .
\label{bellali}\eeq
The function $1/V$ is the product of two factors
which both don't have local maxima
($\Im(\tau_e )$ because it is an harmonic function;
$| a'/\mathcal{W}'|^2$ because it is the squared modulus of an holomorphic function).
So the local maximum, when it exists, comes from a non trivial
interplay between these two different positive  factors.

In order to find metastable vacua one needs to perturb the
$\mathcal{N}=2$ theory with $u$, $u^2$ and $u^3$ operators in the superpotential.
We do not know of any example of  $\mathcal{N}=2$  theory where
metastable vacua are achieved by just adding the $u$ operator;
neither we know about a proof that this can not be achieved.

\subsection{How to generate a metastable vacuum on the moduli space}

\label{ricetta}

Consider a point on the moduli space $u_0$;
then the following
parameterization for the superpotential is introduced:
\beq \mathcal{W}=
\tilde{\mu} \, W=
\tilde{\mu} \left( (u-u_0) +\kappa (u-u_0)^2 + \lambda (u-u_0)^3  \right) 
\, . \label{wii} \eeq
An explicit expression for 
the allowed range of $\kappa$ and $\lambda$
in order to generate   a metastable vacuum in $u_0$
 was found in \cite{pastras2}. In this section
 this calculation is reviewed and some useful notation is introduced.
The potential itself is: 
\beq V=|\tilde{\mu}|^2 \, g^{-1}(u,\bar{u}) \, W'(u) \, \bar{W}'(\bar{u}) \, , \qquad
g=(\im \tau_e (u)) \, a'(u) \, \bar{a}'(\bar{u}) \, ,
\eeq
The first derivative of the potential is computed for $u=u_0$:
\beq \frac{1}{|\tilde{\mu}|^2}  \, 
\frac{\partial V}{\partial u}=
\frac{\partial g^{-1}}{\partial u} \, W'(u) \, \bar{W}'(\bar{u})+
g^{-1} \, W''(u) \, \bar{W}'(\bar{u})
=\frac{\partial g^{-1}}{\partial u}   + 2 \kappa \, g^{-1}   \, .
\eeq
 The condition for generating
  an extremal point at  $u=u_0$ is
\beq \kappa= - \frac{1}{2} \, g \, \frac{\partial g^{-1}}{\partial u}  \, . \label{gg1} \eeq
In order to check
 if this extremal point is a minimum, one needs to calculate
  the second derivatives of $V$:
\beq 
 \frac{1}{|\tilde{\mu}|^2}  
\frac{\partial^2 V}{\partial u^2} =
\frac{\partial^2 g^{-1}}{\partial u^2} \, W' \, \bar{W}' +
2 \frac{\partial g^{-1}}{\partial u} \, W'' \, \bar{W}' + g^{-1} \, W''' \, \bar{W}' \, ,
\eeq
\[  \frac{1}{|\tilde{\mu}|^2}  
\frac{\partial^2 V}{\partial u \partial \bar{u}} =
\frac{\partial^2 g^{-1}}{\partial u \partial \bar{u}} \, W' \, \bar{W}' +
\frac{\partial g^{-1}}{\partial u} \, W' \, \bar{W}'' 
 + \frac{\partial g^{-1}}{\partial \bar{u}} \, W'' \, \bar{W}'+
g^{-1} \, W'' \, \bar{W}'' \,  .
\]
For  $u=u_0$ this reduces to
\beq 
 \frac{1}{|\tilde{\mu}|^2}  
\frac{\partial^2 V}{\partial u^2} =
\frac{\partial^2 g^{-1}}{\partial u^2}
-2 g \left( \frac{\partial g^{-1}}{\partial u} \right)^2 + 6 \lambda \, g^{-1}\, ,
\qquad
  \frac{1}{|\tilde{\mu}|^2}  
\frac{\partial^2 V}{\partial u \partial \bar{u}} =
\frac{\partial^2 g^{-1}}{\partial u \partial \bar{u}} -
g \, \left| \frac{\partial g^{-1}}{\partial u} \right|^2  \,  .
\eeq
A minimum  is obtained if
$ \frac{\partial^2 V}{\partial u \partial \bar{u}} >  \left| \frac{\partial^2 V}{\partial u^2}  \right| $, which gives 
\beq | \lambda-\lambda_0| < \frac{g}{6} \left(
\frac{\partial^2 g^{-1}}{\partial u \partial \bar{u}} 
   -g \, \left| \frac{\partial g^{-1}}{\partial u} \right|^2    \right)
   =r_{\lambda} \, , \label{gg2}
\eeq
where
\beq \lambda_0= \frac{g^2}{3} \left( \frac{\partial g^{-1}}{\partial u} \right)^2- 
\frac{g}{6}  \frac{\partial^2 g^{-1}}{\partial u^2} \, . \label{gg3}
\eeq

The parameterization in terms of $(\kappa,\lambda)$
is useful for the calculation and 
allows to identify the region of the coupling for which
$u_0$ is metastable: it is a ball with radius $r_\lambda$
centered in $\lambda_0$ in the $\lambda$ coordinate, and a point in the $\kappa$ coordinate.
 But on the other hand it is related
 to the physical couplings  by a non-trivial expression involving $u_0$. 
 We find useful to introduce the following parameterization.   
Dropping an irrelevant constant in (\ref{wii}), 
and after an appropriate rescaling, the superpotential 
can be written as
\beq \mathcal{W}=
 \mu \left( u + \alpha
 u^2 + \beta u^3    \right) \, ,  \label{suppe} \eeq  
where
\beq 
\alpha=\frac{\kappa-3 \lambda u_0}
{1-2 \kappa u_0 + 3 \lambda u_0^2} \, , \qquad
\beta=\frac{\lambda}{1-2 \kappa u_0 + 3 \lambda u_0^2} \, .
\eeq 
In the following we will denote as $(\alpha_0,\beta_0)$
the couplings corresponding  to $\lambda=\lambda_0$.
The condition to make $u_0$ metastable is $\lambda=\lambda_0+r_\lambda \epsilon$,
where $\epsilon$ is a complex number with $|\epsilon|<1$.
At the first order in $\delta \lambda=\lambda-\lambda_0$,
which turns out to be a good approximation for the problem, this translates in
\beq
\alpha = \alpha_0 + \delta \alpha \,  \epsilon  \, , \qquad
\delta \alpha=\frac{3(u_0^2 \kappa - u_0) r_\lambda}{(1-2 \kappa u_0 + 3 \lambda u_0^2)^2} \, ,
\eeq
\[ 
\beta = \beta_0 + \delta \beta \,   \epsilon  \, , \qquad
\delta \beta = \frac{(1-2 \kappa u_0 )  r_\lambda}{(1-2 \kappa u_0 + 3 \lambda u_0^2)^2} \, ,
\] 
where the same complex $|\epsilon|<1$ must be chosen for both $(\alpha, \beta)$.

The superpotential (\ref{suppe}) also generates some extra supersymmetric vacua
at the roots of $\mathcal{W}'(u)=0$ :
\beq u_{\pm} = \frac{-\alpha \pm \sqrt{\alpha^2 - 3 \beta}}{3 \beta} \, . \label{extra} \eeq
When the lifetime of the metastable vacuum is considered, also
decays to these extra supersymmetric vacua must be taken into account.

\section{$\mathcal{N}=2$ Super Yang-Mills ($N_f=0$).}

In this case the Seiberg-Witten curve \cite{sw1} is
\beq y^2=(x^2-\Lambda^4)(x-u) \, .\eeq
The singularities on the moduli space,
which correspond to supersymmetric vacua
in the perturbed theory, are at $u_{M,D}=\pm \Lambda^2$.
We set for simplicity the dynamical scale $\Lambda$ to 1.

The functions $(a_D,a)$ can be evaluated by integrating the
Seiberg-Witten differential form on the appropriate cycles
of the curve \cite{sw1}. 
An explicit expression \cite{bilalrev}
in terms of elliptic integrals is:
\beq a(u)
=\frac{ \sqrt{2(1+u)}}{\pi} \,  E \left( \frac{2}{1+u}\right)
\, ,\eeq
\[ a_D(u)
 = \frac{2 i}{\pi} \left( 
(1+u) \, K \left( \frac{1-u}{2}  \right)  -2 E \left(  \frac{1-u}{2}   \right)
\right)  \, .\]
 The following conventions (including also $ \Pi (\nu,k)$, which
 will be useful later) are used in this paper:
\beq 
K(k)=\int_0^{\pi/2} \frac{d \phi}{\sqrt{1-k \sin^2 \phi}} \, ,
\qquad E(k)= \int_0^{\pi/2} \, \sqrt{1-k \sin^2 \phi} \, d \phi \, ,
\label{elliptic}
\eeq
\[ \Pi (\nu,k)=\int_0^{\pi/2} \frac{d \phi}
{(1-\nu \sin^2 \phi) \sqrt{1-k \sin^2 \phi}} \, .
\]

\begin{figure}[h]
\begin{center}
$\begin{array}{c@{\hspace{.2in}}c} \epsfxsize=6.5cm
\epsffile{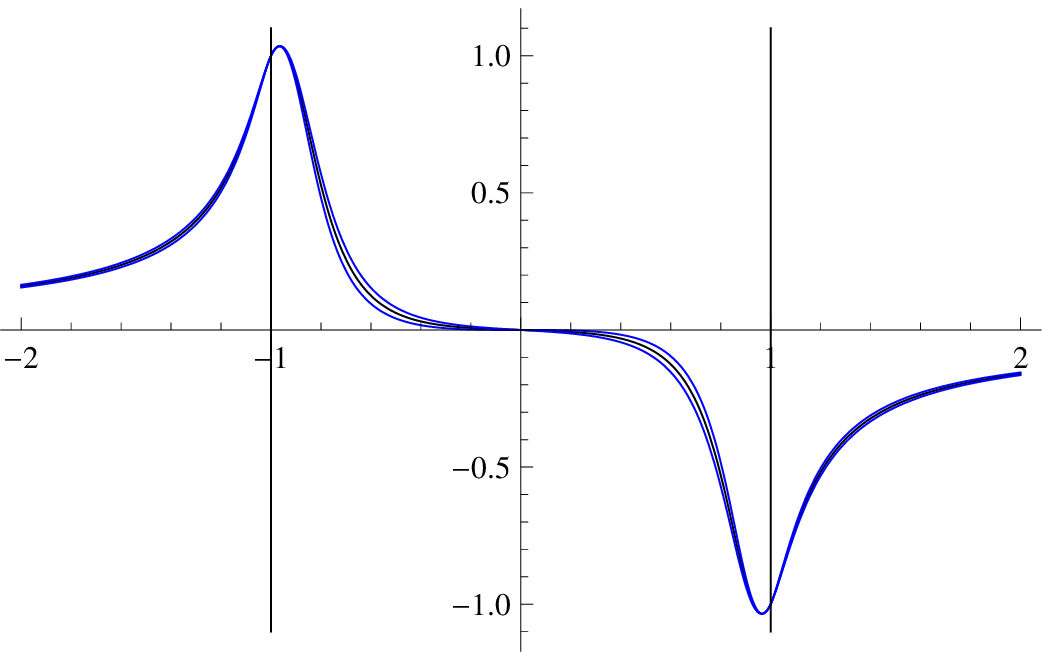} &
    \epsfxsize=6.5cm
    \epsffile{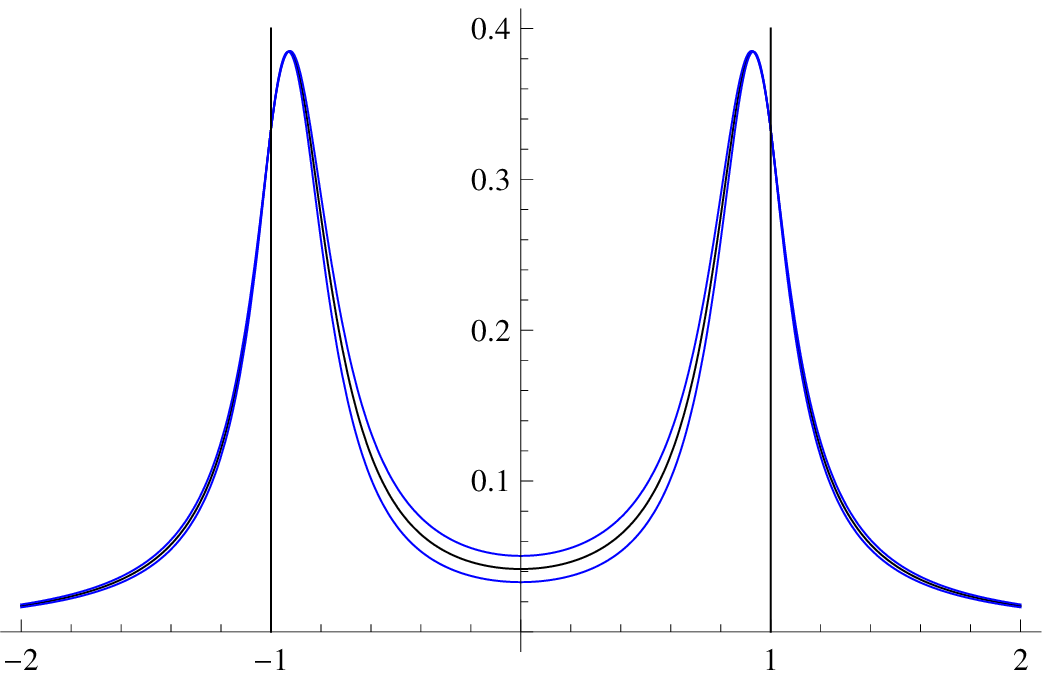}  \end{array}$
\end{center}
\caption{\footnotesize In the black curve, 
values of $\Re \, \alpha_0$ (left)  and  $\Re \, \beta_0$ (right) are plotted
 for $N_f=0$ are as a function of $\Re \, u_0$,
  for $\Im \, u_0 =0$ ($\Im \, \alpha_0, \Im \, \beta_0=0$).
These are the parameters which enter in the superpotential (\ref{suppe}).   
  The curves corresponding to $\alpha_0 \pm \delta \alpha$ and   $\beta_0 \pm \delta \beta$
 are also shown in blue. The vertical lines correspond
 to the location of the supersymmetric vacua.
} \label{fig1}
\end{figure}

\begin{figure}[h]
\begin{center}
$\begin{array}{c@{\hspace{.2in}}c} \epsfxsize=6.5cm
\epsffile{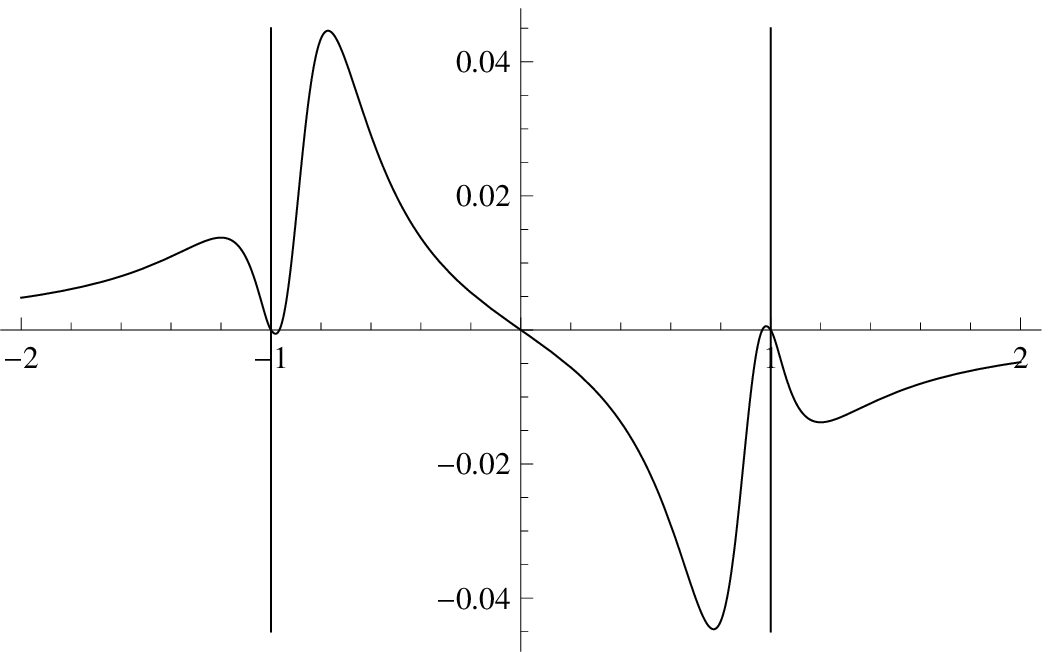} &
    \epsfxsize=6.5cm
    \epsffile{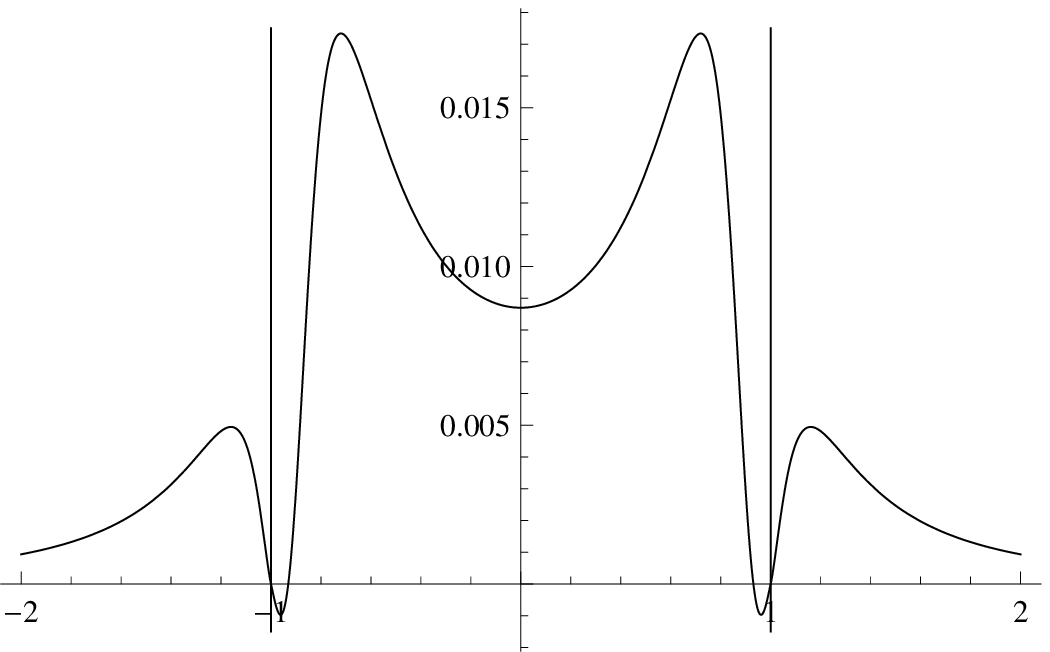}\end{array}$
\end{center}
\caption{\footnotesize
Values of $\delta \alpha$ (left)  and  $\delta \beta$ (right)
 for $N_f=0$,  as a function of $\Re \, u_0$,
 for $\Im \, u_0 =0$. Note that both the functions approach to $0$
 nearby the supersymmetric vacua at $u_0 = \pm \Lambda$.
} \label{fig2}
\end{figure}

\begin{figure}[h]
\begin{center}
\leavevmode
\epsfxsize 6 cm
\epsffile{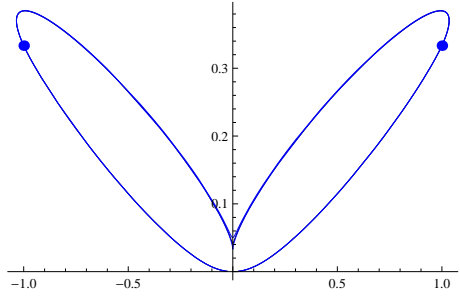}
\end{center}
\caption{\footnotesize  Allowed values of $(\Re \, \alpha,\Re \, \beta)$ in order to get
a metastable vacuum on the real axis of $u_0$ ($\Im \, \alpha,\Im \, \beta=0$) for $N_f=0$.
This correspond to a region between two lines that in the scale of the picture
are almost coincident. The big dots correspond to the limit $u_0 \rightarrow u_{M,D}$,
 in which the would be metastable vacuum does not exist because it coincides
 with a supersymmetric vacuum. }
\label{rabbit}
\end{figure}

\begin{figure}[h]
\begin{center}
\leavevmode
\epsfxsize 6 cm
\epsffile{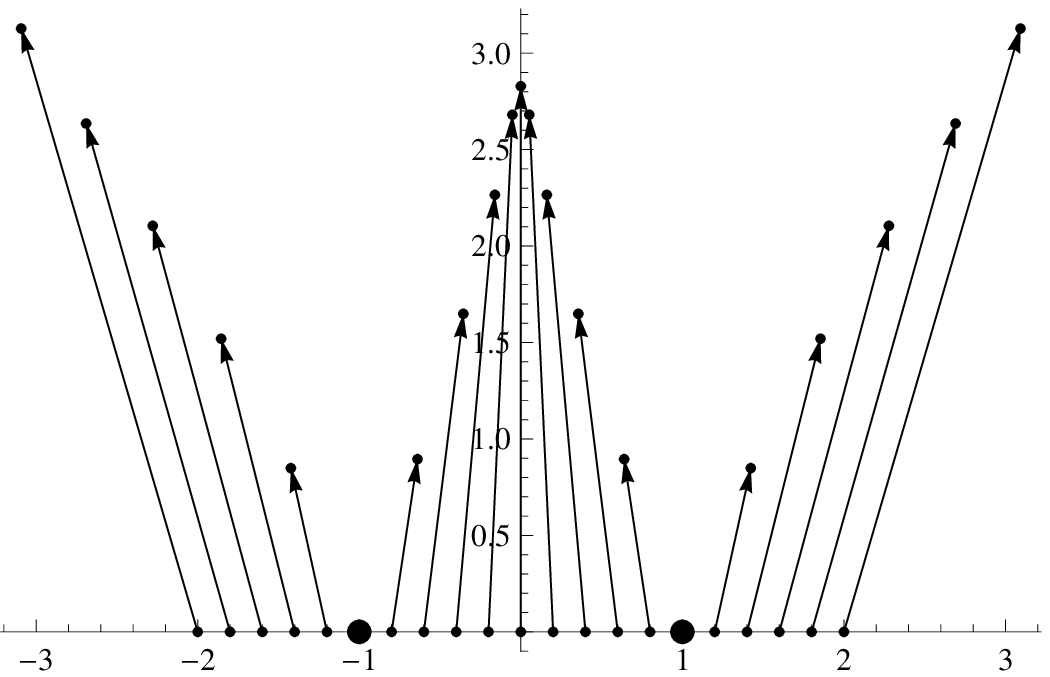}
\end{center}
\caption{\footnotesize  Coordinates of the
 supersymmetric vacuum $(\Re u_{+}, \Im u_{+})$ as a function 
of the metastable vacuum $u_0$, which we take on the real axis. 
The big dots correspond to the supersymmetric vacua $u_{M,D}$;
in the limit $u_0 \rightarrow u_{M,D}$ also $u_{\pm} \rightarrow u_0$. 
Each arrow denotes the $u_+$ vacuum associated with $(\alpha_0,\beta_0)$
of the particular $u_0$. }
\label{arrows0}
\end{figure}

By adding a superpotential $W$ which is a cubic polynomial 
in $u$, one can generate a metastable vacuum at almost every
point $u_0$ of the moduli space, 
 with the exception of the monopole and dyon singularities
at $u_{M,D}$. 
Using the general expressions in section \ref{ricetta},
  the parameters $\alpha_0$,$\beta_0$ can be evaluated
  as a function of the moduli $u_0$. 
The result, for $u_0$ on the real axis, is shown in figure
\ref{fig1}.
 This agrees with the analytical expressions
 for $(\kappa,\lambda_0,r_\lambda)$
  found in  \cite{pastras2}.
In agreement with \cite{OOP,pastras1}, in order to get
a metastable vacuum at the origin of the moduli space, we must choose
\beq \alpha=0 \, , \qquad \beta \approx 0.0417 \pm 0.0087 \, .
\label{origine-nf0}\eeq

Both $\alpha_0$ and $\beta_0$ are regular
nearby the supersymmetric minima at $u_{M,D}$.
The function $(\delta \alpha,\delta \beta)$, which
measure how much we can vary the parameters to
keep $u_0$ metastable, are shown in figure \ref{fig2}.
Note that $(\delta \alpha,\delta \beta)$ go
 both to zero in a linear way as a function of $(u-u_{M,D})$
   when approaching the supersymmetric vacua $u_{M,D}$. 
 
The allowed region of parameters in order to generate
 a metastable vacuum on the real  $u$ axis is shown in figure \ref{rabbit}. 
 The region where the metastable vacuum is less fine-tuned is 
 for $u_0$ nearby the origin,
which corresponds to the parameters in Eq.~(\ref{origine-nf0}).

Using the triangular approximation of \cite{triang},
the tunneling rate of a metastable vacuum to a supersymmetric vacuum 
is proportional to $e^{-S}$, where
\beq S \propto \frac{(\Delta u / \Lambda)^4}{\Delta V} \, , 
\label{tunnel} \eeq
where $\Delta u$ is the distance between the supersymmetric
and the the metastable vacua 
and $\Delta V$ is their difference of potentials.
In order to obtain a long-lived metastable vacuum, the tunneling
rate must be small. This can be achieved by choosing the 
overall constant $\mu$ in Eq. (\ref{suppe}) enough small
(because $\Delta V \propto \mu^2 \Lambda^2$ and $\Delta u$
is independent of $\mu$). 
The tunneling rate  must be checked 
 for the decays to all the supersymmetric vacua 
$u_{susy}=u_{M,D},u_{\pm}$. The vacua $u_{M,D}$
are independent from the deformation parameters; 
on the other hand the vacua $u_+ =u_-^*$ 
are a function of $(\alpha,\beta)$ given by
Eq. (\ref{extra}). 
In figure \ref{arrows0} is it shown $u_+$
as a function of $u_0$; the only limit in which $u_+$
and $u_0$ are almost coincident is for
$u_0 \approx u_{M,D}$. We conclude that the only regime
in which it is problematic to achieve 
a long-lived metastable vacua is for $u_0 \rightarrow u_{M,D}$,
where $u_{M,D}$ are the singularities of the moduli space, 
which correspond to the supersymmetric vacua which do not 
depend on $(\alpha,\beta)$.

With an appropriate rescaling of $\Lambda$, 
the results of this section for the allowed parameter space of $(\alpha,\beta)$
 apply  also to the case with $N_f=2$ massless fundamentals;
 this occurs  because the structure of the singularities on the moduli space and
  the Seiberg-Witten curves are the same.

\section{$N_f=4$ and $\mathcal{N}=2^*$ theories}

As discussed in the introduction, there are no metastable states
in conformal invariant theories; in particular in the $\mathcal{N}=2$
theory with $N_f=2 N_c$ massless fundamentals and in the theory with
$\mathcal{N}=4$ supersymmetries. This argument is no longer valid
in the presence of relevant or irrelevant operators. In this section
we will discuss a situation where both the types of operators are
 present\,\footnote{In \cite{egr} it was found that the particularly interesting
combinations of relevant operators in the $\mathcal{N}=4$ theory
all carry no anomalous dimensions.}.

In the conformal invariant cases 
 the effective coupling $\tau_e$ is a constant
  as a function of the moduli.  
 In the $N_c=2$ case, the functions $(a,a_D)$ are
\beq a= \sqrt{\frac{u}{2}} \, , \qquad a_D=\tau_e a \, 
\qquad {\rm for} \, \, N_f=4 ,\eeq
\[ a= \sqrt{2 u} \, , \qquad a_D=\tau_e a \, 
\qquad {\rm for} \, \,  \mathcal{N}=4 .\]
The moduli space metric is
\beq d \, s^2=(\im \, \tau_e) \, da \, d \bar{a} =
\frac{1}{8}  (u \bar{u})^{-1/2} \, (\im \, \tau_e) \, du \, d \bar{u} . \eeq
From Eqs. (\ref{gg2}) we find that $r_\lambda=0$,
 so it is impossible to stabilize 
any vacuum with a superpotential of the form (\ref{suppe}),
 which is just a function of 
 $u$\,\footnote{The following simpler proof of this statement 
was suggested to us by Zohar Komargodski.
If $\tau_e$ is constant, from Eq.~(\ref{bellali}) it follows that the potential 
 is proportional to the modulus of a holomorphic function and then 
 no classical metastable state with mass gap can exist.
 There could be in principle a pseudo-moduli, 
 but it can be checked that it is not the case.
 This proof can be extended also to the more general case 
 with $N_c>2$, deformed by $W=\Tr \Phi^k$.  }.
Indeed, for the conformal invariant case the proof of \cite{OOP}
does not apply because  the sectional 
curvature of the moduli space is exactly zero.
The situation changes completely when we add
finite masses for some of the hypermultiplets;
then the proof in \cite{OOP} applies and
it is possible to generate a metastable vacuum
at almost every point of the moduli space. 

The expression for $(a,a_D)$
can be found by the Seiberg-Witten curve approach \cite{sw2}.
In the case of the massive $N_f=4$ theory,
with two massless squark hypermultiplet ($m_1=m_2=0$)
and two massive ones with mass $m_3=m_4=m/2$ 
an explicit expression in terms of elliptic integrals was found  in \cite{ferrari} .
Due to the fact that the singularity structure is the same, 
 identical expressions apply also  for
 the $\mathcal{N}=4$ theory deformed by
  mass term $m$ for the adjoint hypermultiplet 
  ($\mathcal{N}=2^*$ theory),
   modulo a trivial  rescaling $(a_D,a)_{\mathcal{N}=2^*}= (a_D, 2 \, a)_{N_f=4}$.
The underlying physics is rather different;
only even instanton corrections contribute in the $N_f=4$
theory while also odd instantons give a non-zero contribution
 in the $\mathcal{N}=2^*$ case. 

In the $N_f=4$ case there are some subtle points in 
the identification between the UV coupling constant and the IR one
which appear in the curve  \cite{dkm1}.
The coupling $\tau_e$
which appear in the SW curve does not
correspond with the $SU(2)$ coupling 
$\tau_{UV}=\theta_{YM}/\pi+8 \pi i /g_{YM}^2$  of the ultraviolet theory.
The expression relating these quantities
gets contributions from an infinite number of instantons \cite{dkm1}:
\beq \tau_e=\tau_{UV} + \frac{i}{\pi} \, \sum_{n=0,2,4,\ldots} a_n q_{UV}^n  \, , \eeq
where $q_{UV}=\exp(i \pi \tau_{UV})$.
For the $\mathcal{N}=2^*$ theory instead 
$\tau_e=\tau_{UV}=\theta_{YM}/(2 \pi)+4 \pi i /g_{YM}^2$
(note the factor of two in the conventions used for the two different cases).

Other subtleties arise both in the $N_f=4$ and in the $\mathcal{N}=2^*$ theories
in the identification of the moduli space coordinate $z$ that appears in the curve
with the operator $u=\Tr \, \Phi^2$ 
of the ultraviolet theory. 
The expression relating this quantities has the following form \cite{dkm1,dkm2}:
\beq z = u  \left( \frac{d \tau}{d \tau_{UV}} \right)^{-1} 
+R \sum_{n=0,2,4,\dots} \alpha_n q^n \, , \qquad
R=\frac{1}{2} \sum_i m_i^2 \, ,\eeq
where different coefficients $\alpha_n$ are needed
in the case of $\mathcal{N}=2^*$ and in the $N_f=4$ case.
This indicates that the operators $z^k$ that
will be introduced in the effective description below 
do not directly correspond to the  operators $(\Tr \, \Phi^2)^k$ in the
UV description but that an unknown 
non-trivial dictionary between the two quantities
is needed. 
 
Following \cite{ferrari}, we introduce the parameter
\beq
\tilde{u}=z +\frac{1}{8} m^2 E_1(\tau_e) \, ,
\eeq
where $E_1$ is a function that will be defined in the next paragraph.
In this way the parameter $\tilde{u}$ is identified
with the physical parameter $u=\Tr \, \Phi^2$ parameter under the renormalization
group flow  from the $N_f=4$ to the $N_f=2$ theory 
(or from the $\mathcal{N}=4$  to the Super-Yang-Mills $N_f=0$).
The variable $\tilde{u}$ then can be identified with $u$ at least in the decoupling limit
$m \rightarrow \infty$, $g_{YM} \rightarrow 0$ with $\Lambda \propto m e^{-1/g_{YM}^2}$ 
fixed.

The Seiberg-Witten curve \cite{sw2} is:
\beq y^2=4 (x-e_1 )(x-e_2) (x-e_3) \, ,\eeq
where 
\beq e_1=0  \, , \qquad
 e_3=(E_2(\tau_e) -E_1(\tau_e)) z + \frac{1}{4} m^2 (E_2^2(\tau_e)-E_1^2(\tau_e)) \, ,
 \label{eee} \eeq
\[ e_2=(E_3(\tau_e) -E_1(\tau_e)) z + \frac{1}{4} m^2 (E_3^2(\tau_e)-E_1^2(\tau)) \, .\]
The functions $E_j(\tau_e)$ are defined in term of standard 
$\theta_{1,2,3}(\tau_e)$ functions:
\beq E_1(\tau_e)= \frac{\theta_2^4 + \theta_3^4}{3} \, , \qquad
E_2(\tau_e)=-\frac{\theta_1^4+\theta_3^4}{3}  \, , \qquad
E_3(\tau_e)= \frac{\theta_1^4-\theta_2^4}{3} \, .
\eeq
The following three values of $z$ correspond to 
singularities of the moduli space metric, where extra massless
(electric or magnetic) degrees of freedom  are present:
\beq z_j=\frac{m^2}{4} E_j(\tau_e) \, , \qquad
j=1,2,3 \, .\eeq
These values correspond to supersymmetric vacua;
we denote the corresponding values of $\tilde{u}$
as $(\tilde{u}_{s1},\tilde{u}_{s2},\tilde{u}_{s3})$.

It is  useful to introduce the variables:
\beq  k^2=\frac{e_2-e_3}{e_1-e_3} \, , \qquad
k'^2=1-k^2=\frac{e_2-e_1}{e_3-e_1} \, .  \label{kakappa} \eeq
The general strategy is to reduce the computation of $a,a_D$
to the following three elliptic integrals, which can be expressed 
in term of standard special functions:
\beq 
I_1^j=\oint_{\gamma_j} \frac{dx}{y} \, , \qquad
I_2^j=\oint_{\gamma_j} \frac{x \, dx}{y} \, , \qquad
I_3^j(c)=\oint_{\gamma_j} \frac{dx}{(x-c) \, y} \, .
\eeq
The explicit expressions \cite{ferrari,bf2} are:
\beq I_1^1 = \frac{2}{\sqrt{e_1-e_3}}  K(k^2) \, , \qquad
  I_2^1 = \frac{2}{\sqrt{e_1-e_3}} ( e_1 K(k^2)   +(e_3-e_1) E(k^2)) \, , \eeq 
\[  I_3^1(c)=\frac{2}{(e_1-e_3)^{3/2}} \left(
\frac{1}{1-\tilde{c} + k'} K(k^2)+ \frac{4 k'}{1+k'} \frac{1}{(1-\tilde{c})^2 - k'^2}
\Pi \left(\nu, \left(\frac{1-k'}{1+k'}\right)^2 \right)
\right) \]
where
\[ \tilde{c}=\frac{c-e_3}{e_1-e_3} \, , \qquad
\nu=\left( \frac{1-\tilde{c} +k'}{1-\tilde{c}-k' } \right)^2  
\left(\frac{1-k'}{1+k'}\right)^2 \, ,\]
and $I_j^2$ can be obtained from $ I_j^1$  by exchanging $e_1$ and $e_3$
(this exchanges  $k$ and $k'$).
The functions $K,E, \Pi$ are standard elliptic integrals,
defined with the conventions in Eq. (\ref{elliptic}).

Then the explicit expression for $(a,a_D)$ can be evaluated,
by integrating the SW differential; the result is
\beq 
a(z)=\frac{\sqrt{2}}{ \pi} (z-z_1)
\left( I_1^1 -\frac{m^2}{4} \theta_2^4 \theta_3^4 \, I_3^1  \left( \frac{m^2}{4} \theta_2^4 \theta_3^4 \right) \, \right) \, ,
\eeq
\[ 
a_D(z)=\frac{\sqrt{2}}{ \pi} (z-z_1)
\left( I_1^2 -\frac{m^2}{4} \theta_2^4 \theta_3^4 \, I_3^2  \left( \frac{m^2}{4} \theta_2^4 \theta_3^4 \right) \, \right) \, .
\]
The value of the low energy coupling $\tau_e$ is
\beq \tau_e=\frac{d \, a_D}{d a} =\frac{I^2_1}{I^1_1} \, .\eeq

\begin{figure}[h]
\begin{center}
$\begin{array}{c@{\hspace{.2in}}c} \epsfxsize=6cm
\epsffile{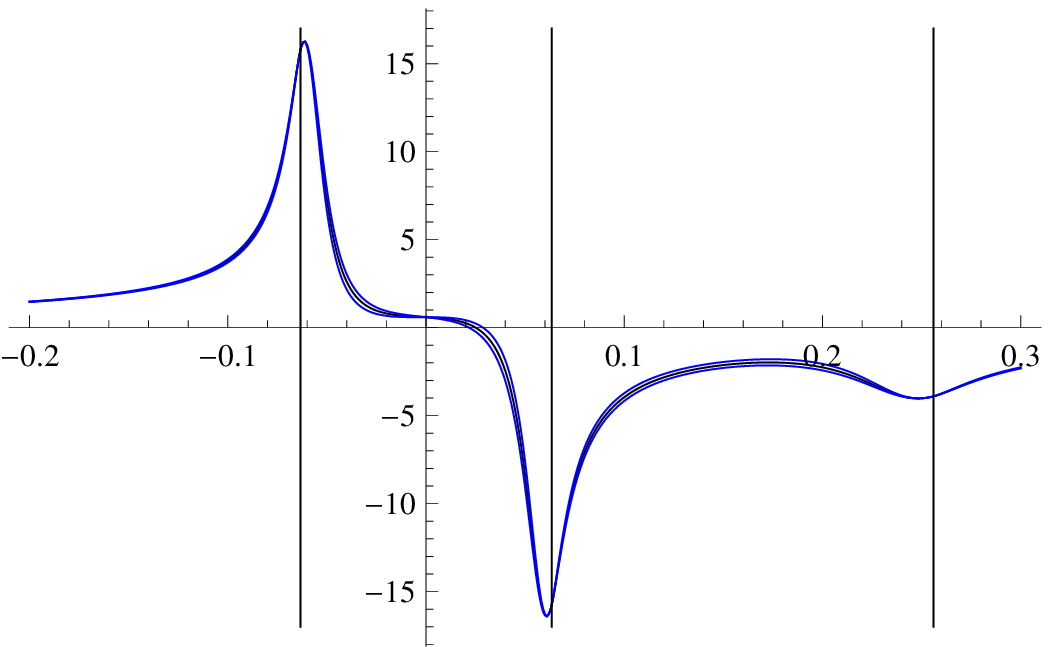} &
     \epsfxsize=6cm
    \epsffile{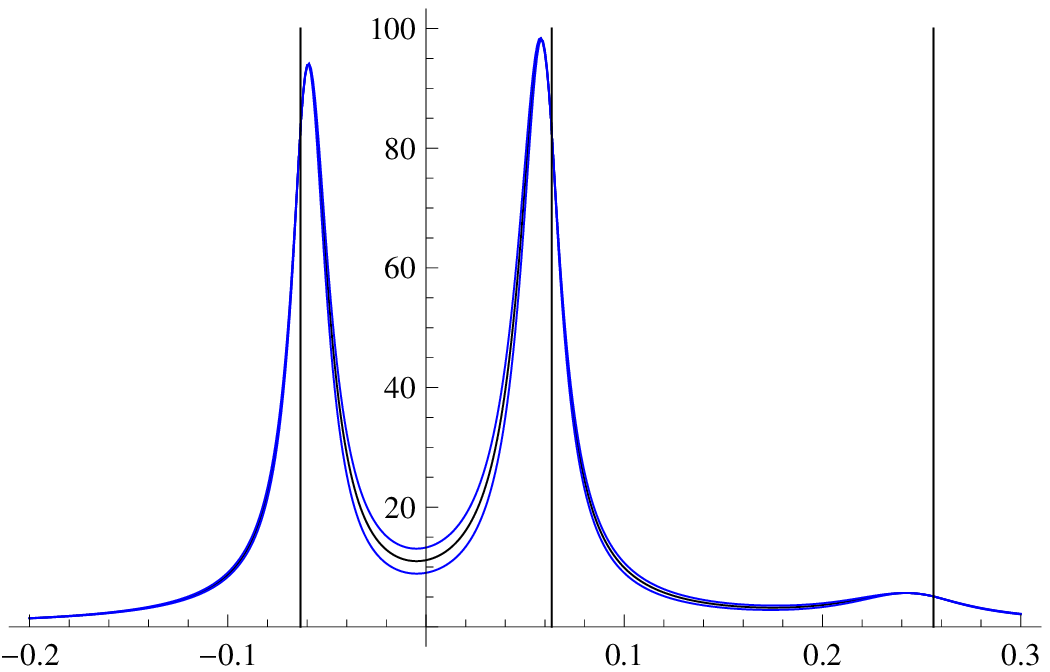}
\end{array}$
\end{center}
\caption{\footnotesize
Left:  $\Re \, \tilde{\alpha}_0$, Right: $\Re \, \tilde{\beta}_0$  for $\tilde{u}_0$ real 
 ($\Im \, \tilde{\alpha}_0,\Im \, \tilde{\beta}_0=0$) for $N_f=4$ 
and $\tau_e= 1.1 i$ (the units are set by $m=1$). The same plot
applies to the $\mathcal{N}=2^*$ theory.
} \label{nf4a}
\end{figure}

\begin{figure}[h]
\begin{center}
$\begin{array}{c@{\hspace{.2in}}c} \epsfxsize=6cm
\epsffile{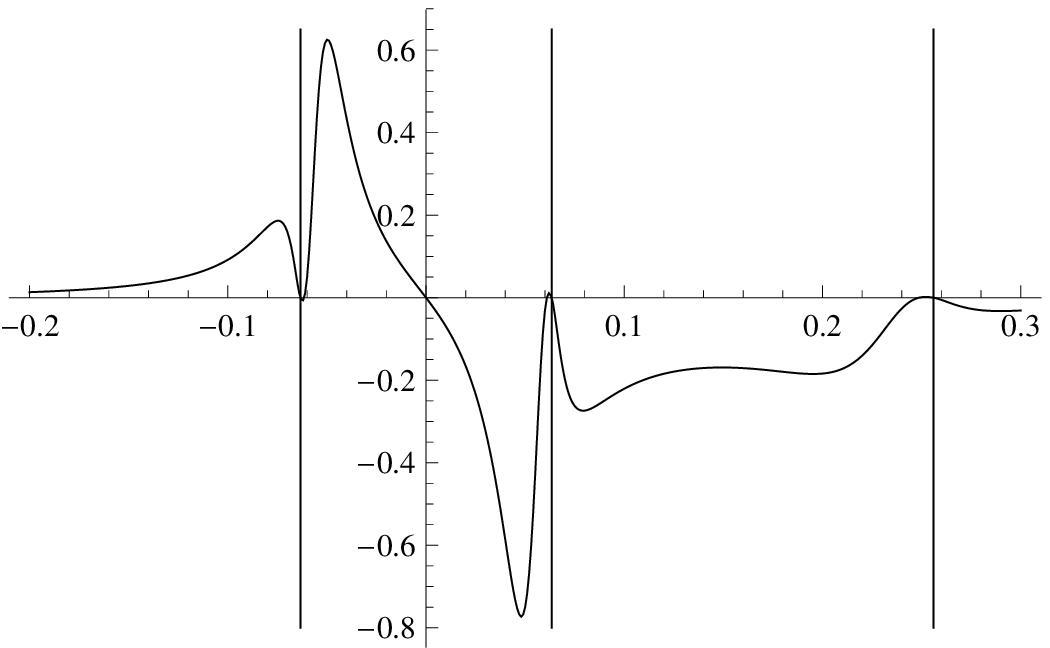} &
     \epsfxsize=6cm
    \epsffile{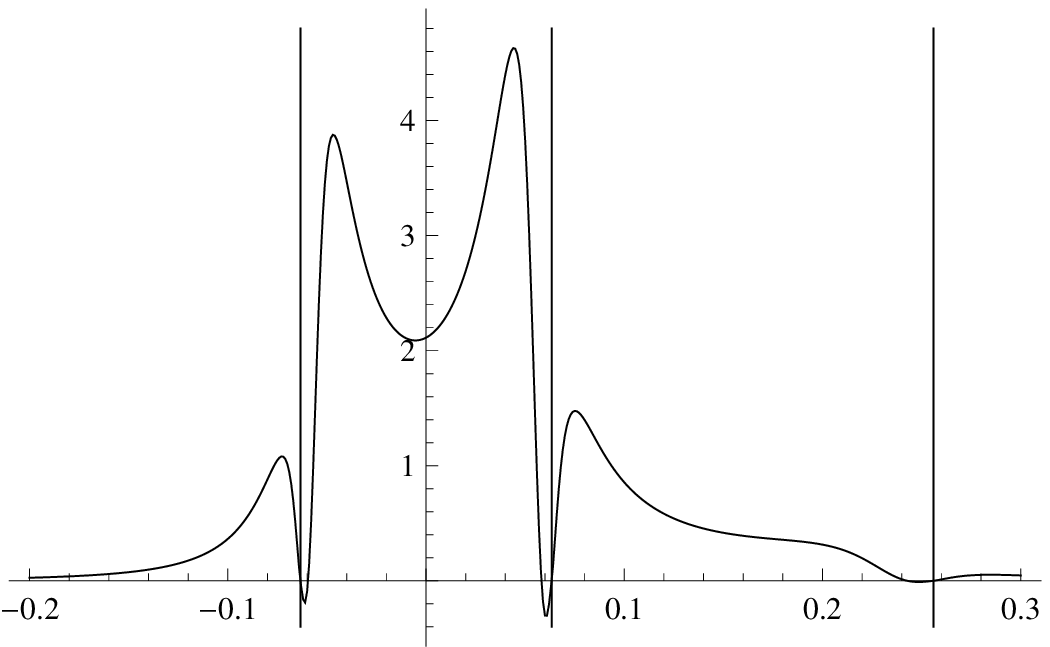}
\end{array}$
\end{center}
\caption{\footnotesize
Left:  $\delta \tilde{\alpha}$, Right: $\delta \tilde{\beta}$  for $\tilde{u}_0$ real
($N_f=4$, $\tau_e=1.1 i$ and $m=1$).
} \label{nf4b}
\end{figure}

\begin{figure}[h]
\begin{center}
\leavevmode
\epsfxsize 6 cm
\epsffile{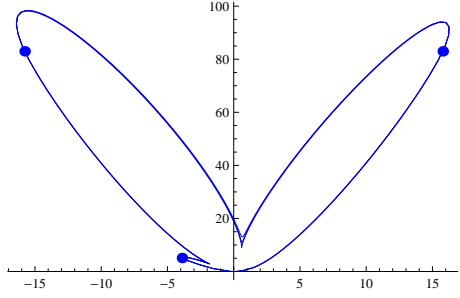}
\end{center}
\caption{\footnotesize  Allowed values of $(\Re \, \tilde{\alpha},\Re \, \tilde{\beta})$ in order to get a metastable vacuum on the real axis of $\tilde{u}_0$ (with $\Im \, \tilde{\alpha},\Im \, \tilde{\beta}=0$), 
for $N_f=4$, $\tau_e= 1.1 i$, $m=1$. }
\label{nf4c}
\end{figure}

Using the general expression in section \ref{ricetta},
we can now study the allowed region of the parameters $(\tilde{\alpha},\tilde{\beta})$ in
\[ \mathcal{W}= \mu (\tilde{u} + \tilde{\alpha} \tilde{u}^2+ \tilde{\beta} \tilde{u}^3 ) \, \]
in order to generate a metastable vacuum in $\tilde{u}_0$.
The variable $\tilde{u}$ can be identified with $u =\Tr \Phi^2$ just in a decoupling limit
where we recover the massless $N_f=2$ case for the massive $N_f=4$ theory
and the $N_f=0$ case for the $\mathcal{N}=2^*$ theory.
The full expression for $\tilde{u}$ as a function of $u$
is unknown in both the cases.
Plots for the quantities $(\tilde{\alpha}_0,\tilde{\beta}_0,\delta \tilde{\alpha}, \delta \tilde{\beta})$,
for the value $\tau=1.1 i$,
are shown in figure \ref{nf4a}, \ref{nf4b}. 
The allowed region of parameters
in order to generate 
a metastable vacuum on the real $\tilde{u_0}$ axis is shown in
 figure \ref{nf4c}.

\begin{figure}[h]
\begin{center}
\leavevmode
\epsfxsize 6 cm
\epsffile{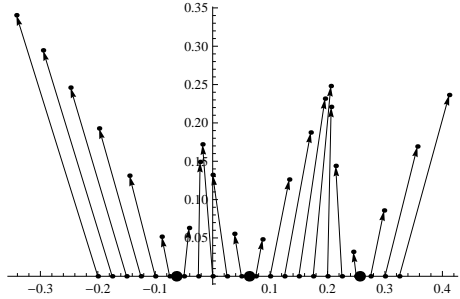}
\end{center}
\caption{\footnotesize   Coordinates of the
 supersymmetric vacuum $(\Re \tilde{u}_{+}, \Im \tilde{u}_{+})$ as a function 
  the metastable vacuum $\tilde{u}_0$,  
for $N_f=4$, $\tau_e= 1.1 i$, $m=1$. }
\label{nf4d}
\end{figure}

The tunneling
rate can be estimated by Eq. (\ref{tunnel}).
In order to check that the metastable 
vacuum is long-lived, we must check 
the decays to all the supersymmetric vacua 
$\tilde{u}_{susy}=\tilde{u}_{s1},\tilde{u}_{s2},\tilde{u}_{s3},\tilde{u}_{\pm}$.
 In figure \ref{nf4d}  the supersymmetric vacuum $\tilde{u}_+$ is  shown 
as a function of $\tilde{u}_0$;  $\tilde{u}_+$
and $\tilde{u}_0$ are almost coincident just for
$\tilde{u}_0 \rightarrow \tilde{u}_{s1},\tilde{u}_{s2},\tilde{u}_{s3}$.
The only limit in which it is problematic to achieve 
a long-lived metastable vacua is when $\tilde{u}_0$
is chosen very nearby to these values.

It is interesting that in the case with fundamental hypermultiplets 
it is possible to obtain metastable vacua also in another limit.
Let us start with $\mathcal{N}=2$ $SU(N_c)$ gauge theory
with $N_f=2 N_c$ fundamentals. 
In $\mathcal{N}=1$ language, the field content  is
given by a vector superfield, and adjoint chiral superfield $\Phi$,
and $N_f$ fundamentals and anti-fundamentals $Q$ and $\tilde{Q}$.
The superpotential reads:
\beq W=\sum_{i=1 \ldots N_f} \tilde{Q}_i \Phi Q_i \, .\eeq
Let us then fix an integer $\tilde{N}_f$ with
$ N_c < \tilde{N}_f < 3/2 N_c $.
The following mass terms
are then introduced in the superpotential:
\beq \Delta W= M \Phi^2+ \sum_{i=\tilde{N}_f+1 \ldots N_f} M Q \tilde{Q}+
\sum_{i=1 \ldots \tilde{N}_f} m Q \tilde{Q} \, .
\eeq
Then we consider the limit  $g_{YM} \rightarrow 0$.
In this limit, at the scale $M$ some of the fields of the theory
decouple  and do not contribute any more to
the $\beta$ function for the gauge coupling.
In the far infrared, the theory reduces to 
$\mathcal{N}=1$ SQCD with dynamical scale 
$ \Lambda \approx M e^{-1/g_{YM}^2}$.
The mass term $m$ is then chosen in such a way that
$ m << \Lambda$. 
We can now embed the ISS 
model \cite{ISS} in the infrared of the theory.
The range $ N_c < \tilde{N}_f < 3/2 N_c $ is needed in order
for the Seiberg dual to be in the free magnetic phase.
The mass term $m$ is also 
needed in order to have metastable supersymmetry breaking.
This metastable vacuum is rather different 
from the ones that are found on the 
Coulomb branch in the limit of small perturbation
from the $\mathcal{N}=2$ limit. It is not known
if these two kinds of vacua can be related by 
a continuous change of the parameters.

\section{An example with a
conformal point: the $N_f=2$ theory}

In this section we discuss the case of $N_f=2$ massive fundamentals,
which is interesting because for a critical value of the
hypermultiplet mass ($m_1=m_2=\Lambda/2$) an infrared 
Argyres-Douglas fixed point \cite{adapsw} exists.

The Seiberg-Witten curve \cite{sw2} in this case is:
\beq y^2=x^2 (x-u) -\frac{\Lambda^4}{64} (x-u)
+\frac{\Lambda^2}{4} m_1 m_2 x-\frac{\Lambda^4}{64}(m_1^2+m_2^2) \, .\eeq
In this section we  set $m_1=m_2=m$.
The singular points of the moduli space of vacua are at
\beq u_{s1}=-\frac{\Lambda^2}{8} -\Lambda m \, , \qquad
u_{s2}= -\frac{\Lambda^2}{8} +\Lambda m \, , \qquad
u_{s3}=m^2+\frac{\Lambda^2}{8} \, . \label{ussusy}
\eeq
These values correspond to supersymmetric vacua.

The roots of the polynomial which defines the cubic are
\beq e_1=\frac{u}{6}-\frac{\Lambda^2}{16}
+\frac{1}{2} \sqrt{u+\frac{\Lambda^2}{8} + \Lambda m}
 \sqrt{u+\frac{\Lambda^2}{8} - \Lambda m} \, , \label{eeenf2}
\eeq
\[ e_2 = -\frac{u}{3} + \frac{\Lambda^2}{8} \, ,\]
\[ e_3=\frac{u}{6}-\frac{\Lambda^2}{16}
-\frac{1}{2} \sqrt{u+\frac{\Lambda^2}{8} + \Lambda m}
 \sqrt{u+\frac{\Lambda^2}{8} - \Lambda m} \, ,\]
where a translation in such a way that $\sum_i e_i=0$ is done for
convenience.
We can then define the expressions
  for $I_i^j$ in the same 
way as for $N_f=4$, using the new $e_j$
given in Eq. (\ref{eeenf2}) and $k,k'$ defined as in Eq. (\ref{kakappa}) in term of the new $e_j$.
 These expressions are now function of 
the moduli space coordinate $u$ instead that of $z$ as in the $N_f=4$ case.

\begin{figure}[h]
\begin{center}
$\begin{array}{c@{\hspace{.2in}}c} \epsfxsize=6cm
\epsffile{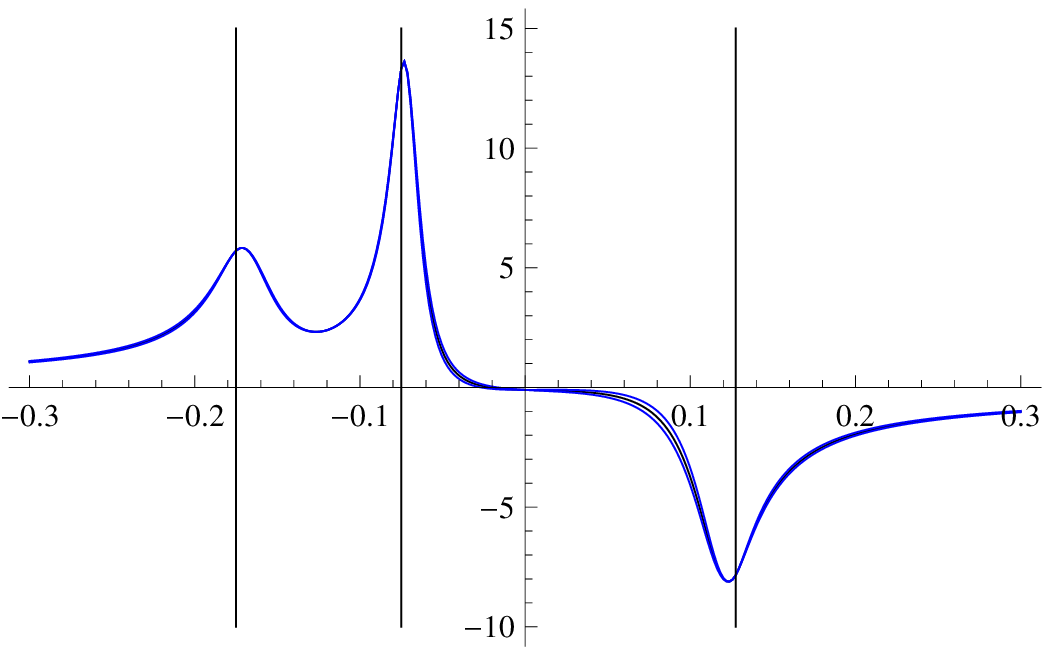} &
     \epsfxsize=6cm
    \epsffile{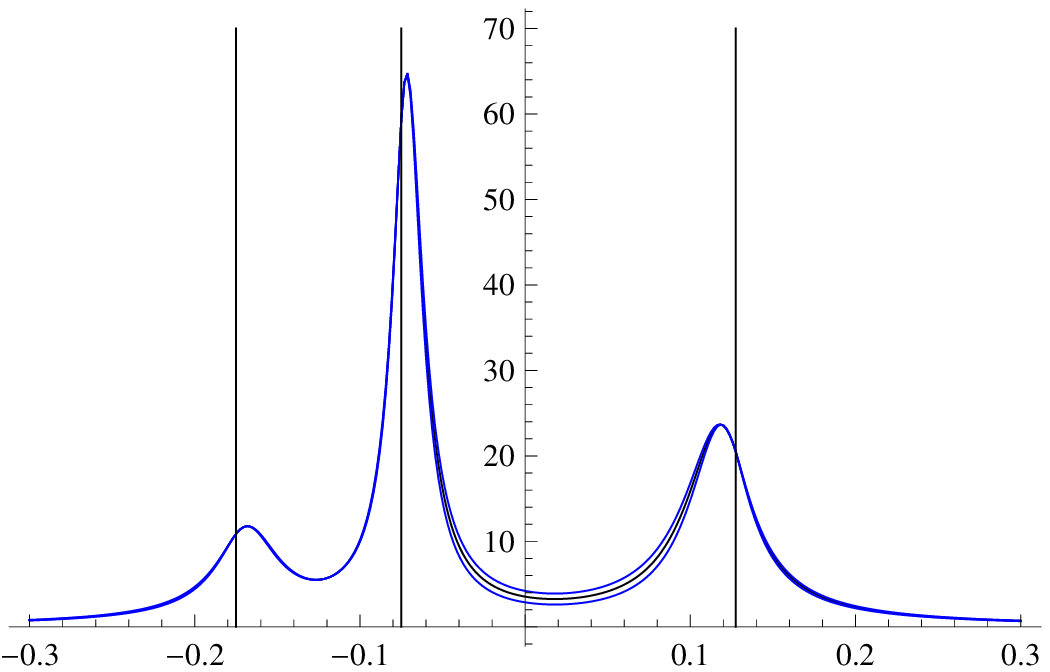}
\end{array}$
\end{center}
\caption{\footnotesize
Left:  $\alpha_0$, Right: $ \beta_0$  
on the real $u_0$ axis for $N_f=2$ 
and $m=0.05 \Lambda$. The units are fixed by $\Lambda=1$.
} \label{nf2a}
\end{figure}

\begin{figure}[h]
\begin{center}
$\begin{array}{c@{\hspace{.2in}}c} \epsfxsize=6.5cm
\epsffile{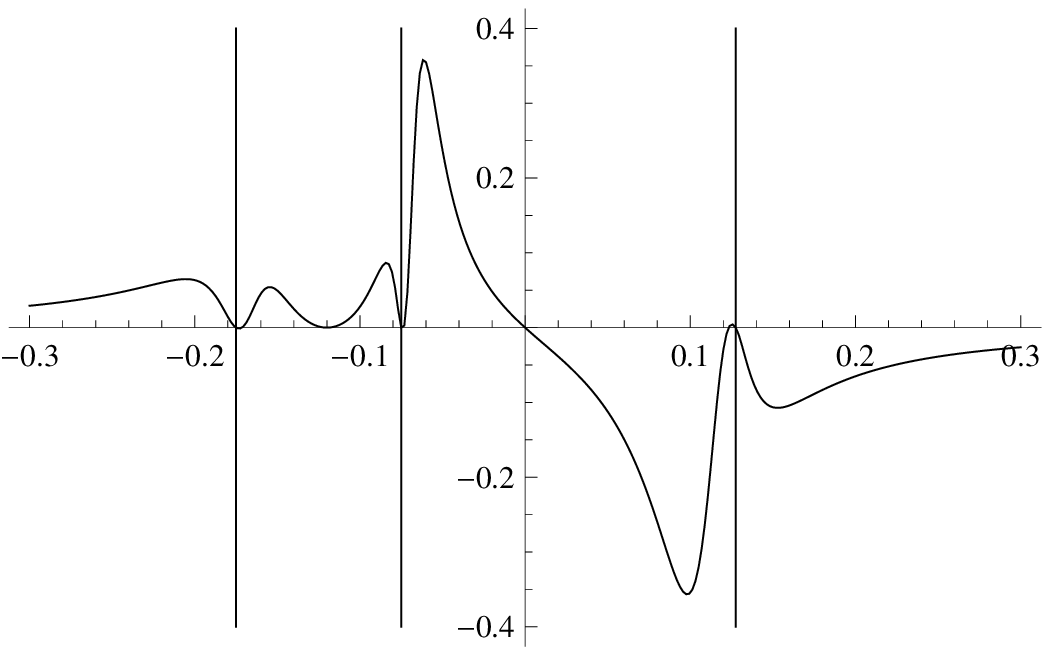} &
    \epsfxsize=6.5cm
    \epsffile{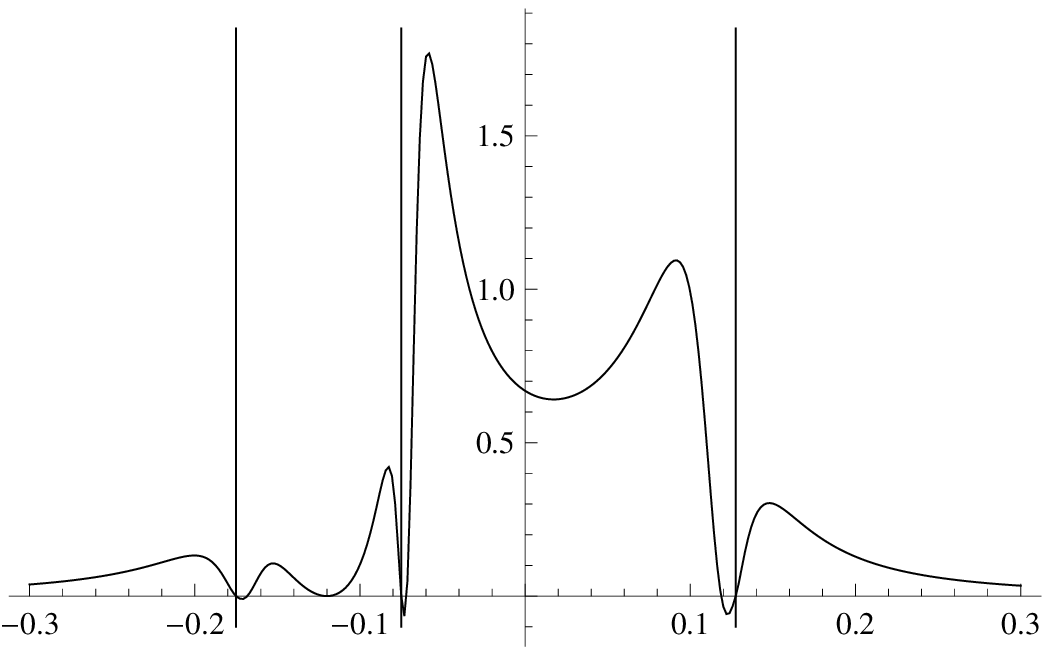}\end{array}$
\end{center}
\caption{\footnotesize
Values of $\delta \alpha$ (left)  and  $\delta \beta$ (right)
 for $N_f=2$ for $m=0.05 \Lambda$ on the real $u_0$ axis.
} \label{nf2b}
\end{figure}

\begin{figure}[h]
\begin{center}
$\begin{array}{c@{\hspace{.2in}}c} \epsfxsize=6cm
\epsffile{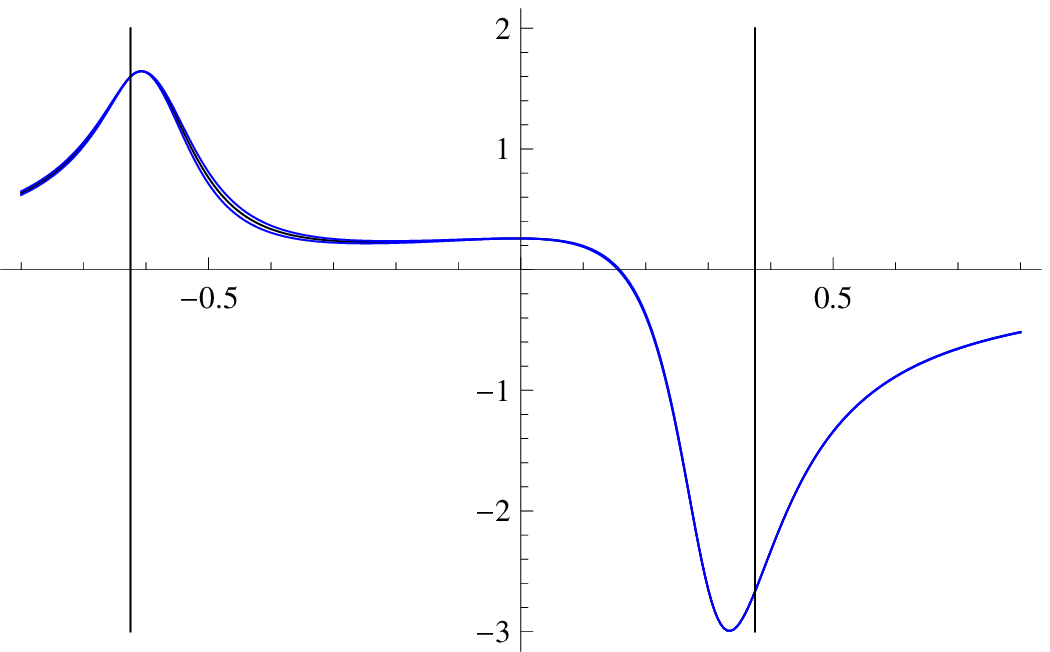} &
     \epsfxsize=6cm
    \epsffile{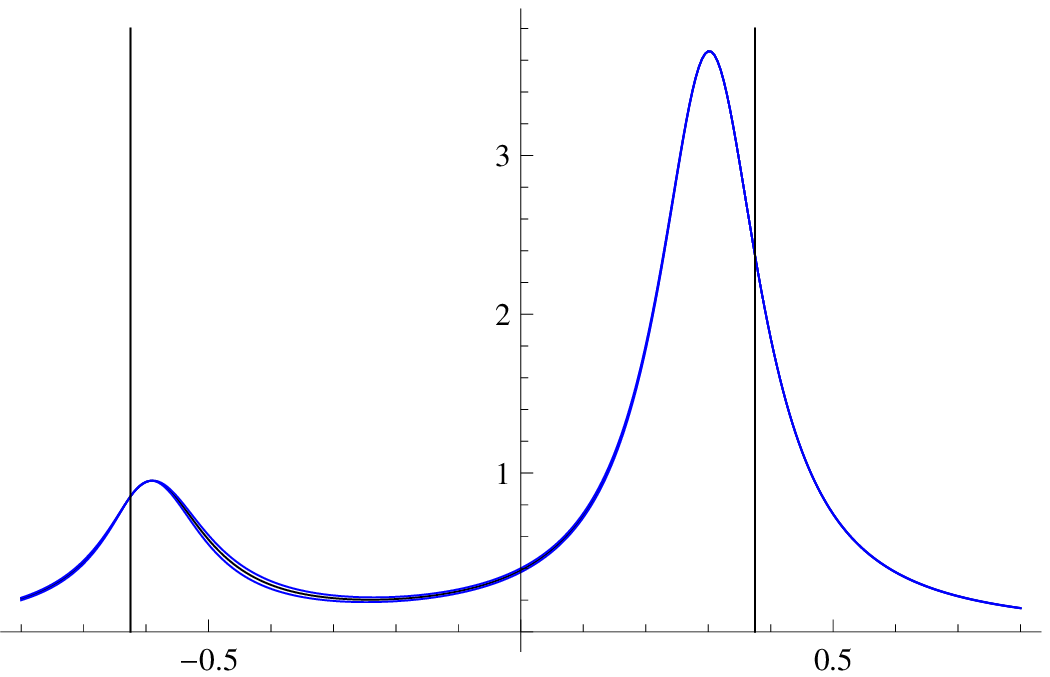}
\end{array}$
\end{center}
\caption{\footnotesize
Left:  $\alpha_0$, Right: $ \beta_0$  
on the real $u_0$ axis for $N_f=2$ 
and $m=m_c=0.5 \Lambda$ (for this value there is an Argyres-Douglas point).
 The AD fixed point is at $u_{AD}=0.375 \Lambda^2$.
} \label{nf2c}
\end{figure}

\begin{figure}[h]
\begin{center}
$\begin{array}{c@{\hspace{.2in}}c} \epsfxsize=6.5cm
\epsffile{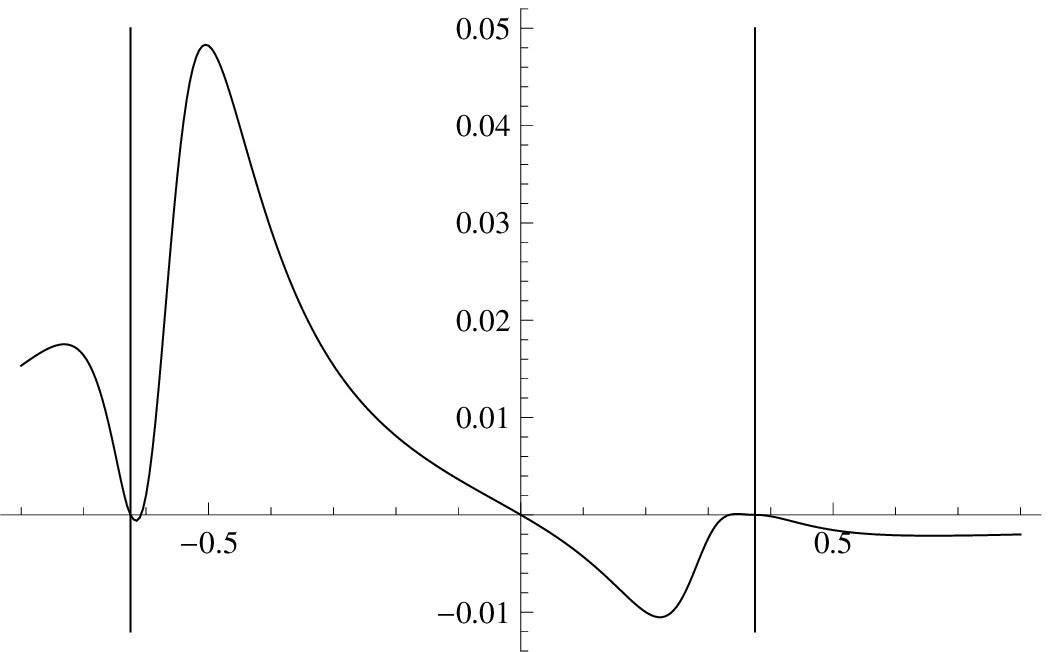} &
    \epsfxsize=6.5cm
    \epsffile{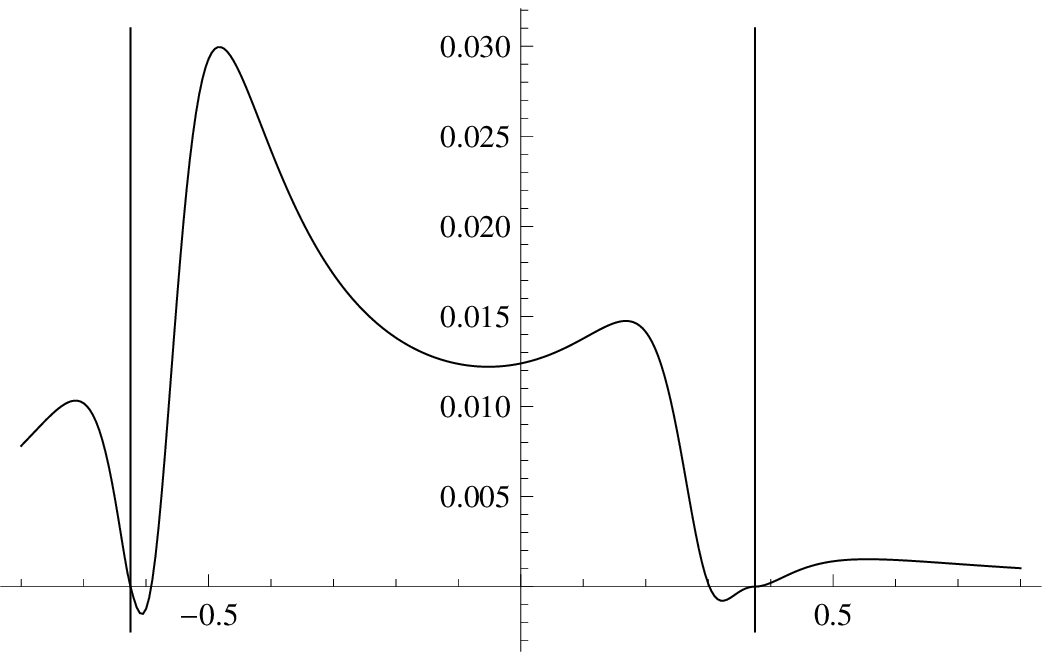}\end{array}$
\end{center}
\caption{\footnotesize
Values of $\delta \alpha$ (left)  and  $\delta \beta$ (right)
 for $N_f=2$ for $m=m_c$,  on the real $u_0$ axis.
} \label{nf2d}
\end{figure}

\begin{figure}[h]
\begin{center}
$\begin{array}{c@{\hspace{.2in}}c} \epsfxsize=6.5cm
\epsffile{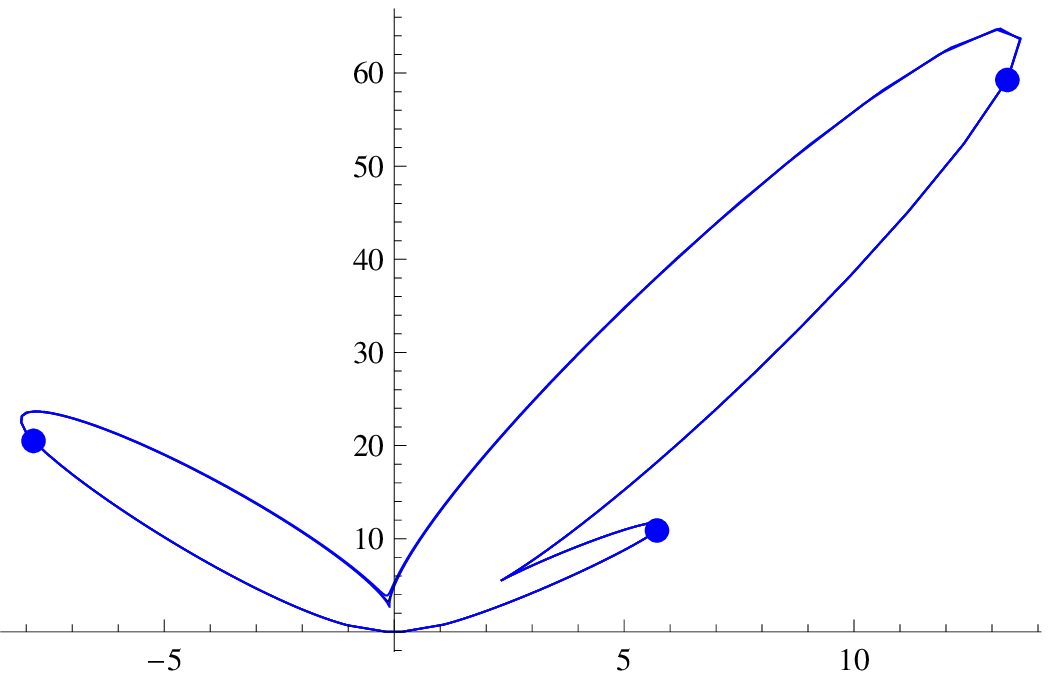} &
     \epsfxsize=6.5cm
    \epsffile{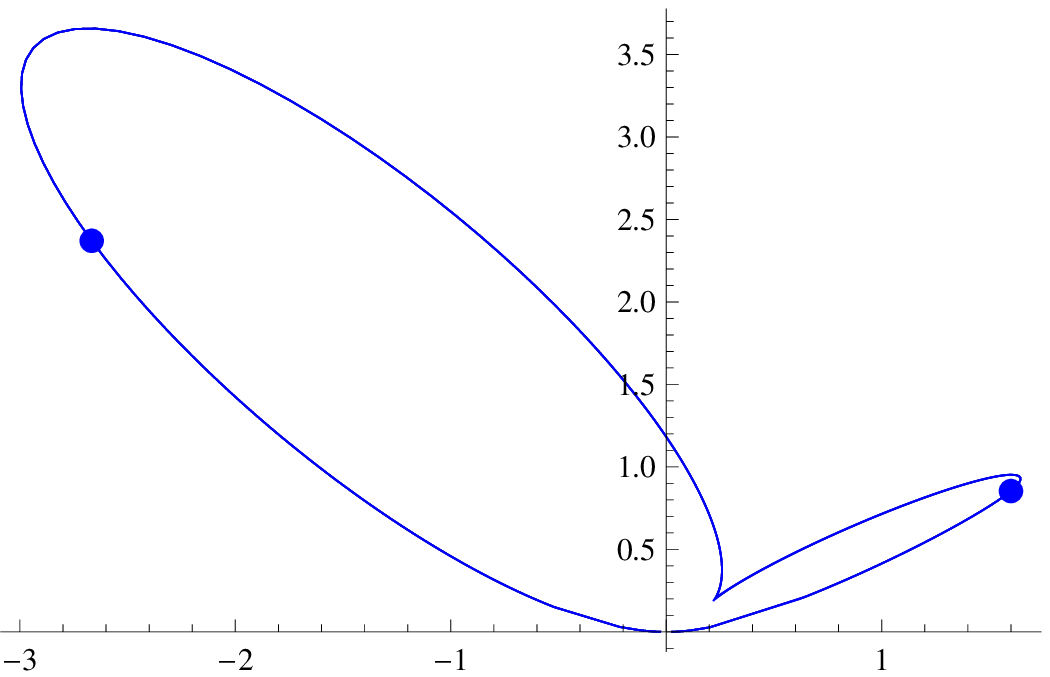}
\end{array}$
\end{center}
\caption{\footnotesize
Allowed values of $(\Re \, \alpha,\Re \, \beta)$ in order to get
a metastable vacuum on the real axis of $u_0$ (with $\Im \, \alpha,\Im \, \beta=0$) for 
$N_f=2$, with $m=0.05 \Lambda$ (left) and  $m=m_c=0.5 \Lambda$ (right).
} \label{nf2bb}
\end{figure}

 An explicit expression for $(a,a_D)$ was computed in \cite{bf2}:
\beq
a=\frac{\sqrt{2}}{4 \pi}
\left( \frac{4}{3} \, u I_1^1 - 2 I_2^1
 -\frac{\Lambda^2}{2} m^2 I_3^1 \left( -\frac{\Lambda^2}{8} -\frac{u}{3} \right)
\right) + \frac{m}{\sqrt{2}} \, ,
\eeq
\[ a_D = 
\frac{\sqrt{2}}{4 \pi}
\left( \frac{4}{3} \, u I_1^2 - 2 I_2^2
 -\frac{\Lambda^2}{2} m^2 I_3^2 \left( -\frac{\Lambda^2}{8} -\frac{u}{3} \right)
\right) \, .
\]
Using the general expressions in section \ref{ricetta},
we can then compute the allowed region of the parameters $(\alpha,\beta)$.
The values of $(\alpha_0,\beta_0,\delta \alpha, \delta \beta)$
in the case of $m=0.05 \Lambda$ on the real $u_0$ axis are shown
in figures \ref{nf2a}  and  \ref{nf2b}. Similar plots for the critical
mass $m_c=0.5 \Lambda$  are shown in figure  \ref{nf2c}  and  \ref{nf2d};
the allowed region of parameters for the two masses choices is shown in figure 
 \ref{nf2bb}.

For generic $m$ there are three singularities on the moduli space (see Eq. (\ref{ussusy}));
for $m_c=\Lambda/2$, two of these singularities collide
and an Argyres-Douglas point appears \cite{adapsw}, corresponding to
a non-trivial interacting conformal fixed point in the IR.
The AD fixed point is at $u_{AD}=0.375 \Lambda^2$.
For $m>m_c$ one of the singularities is at weak coupling and corresponds to 
 massless electric degrees of freedoms; 
the other two singularities are in the strong coupling region and correspond to the
massless monopole and dyon  points of the $N_f=0$ case.
For $m<m_c$ all the singularities are at strong coupling.

\begin{figure}[h]
\begin{center}
$\begin{array}{c@{\hspace{.2in}}c} \epsfxsize=6cm
\epsffile{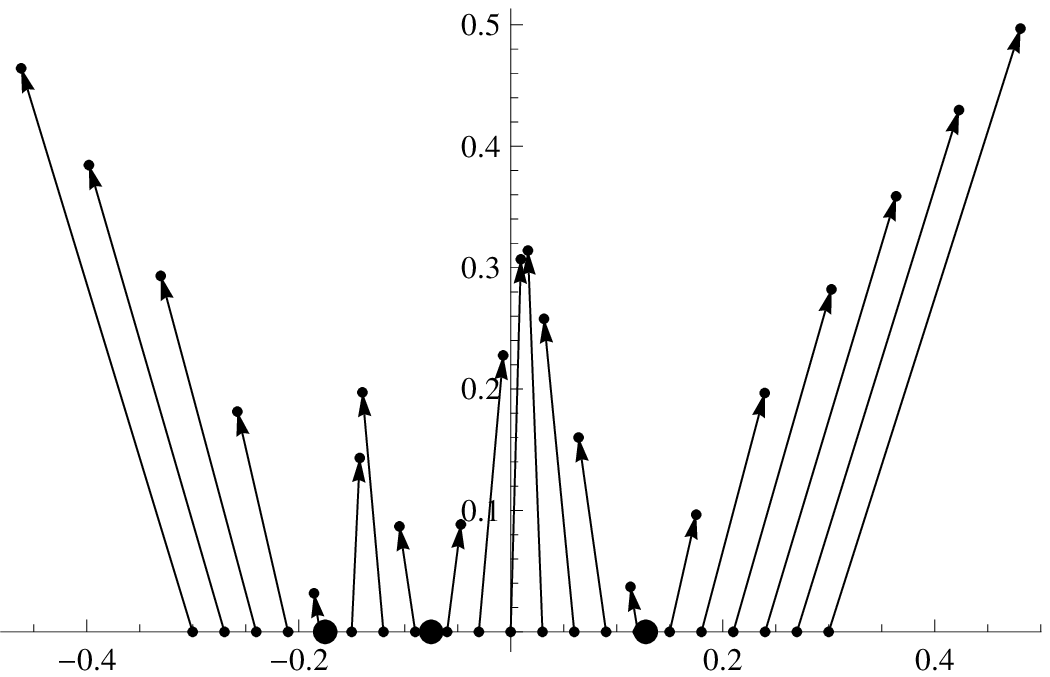} &
     \epsfxsize=6cm
    \epsffile{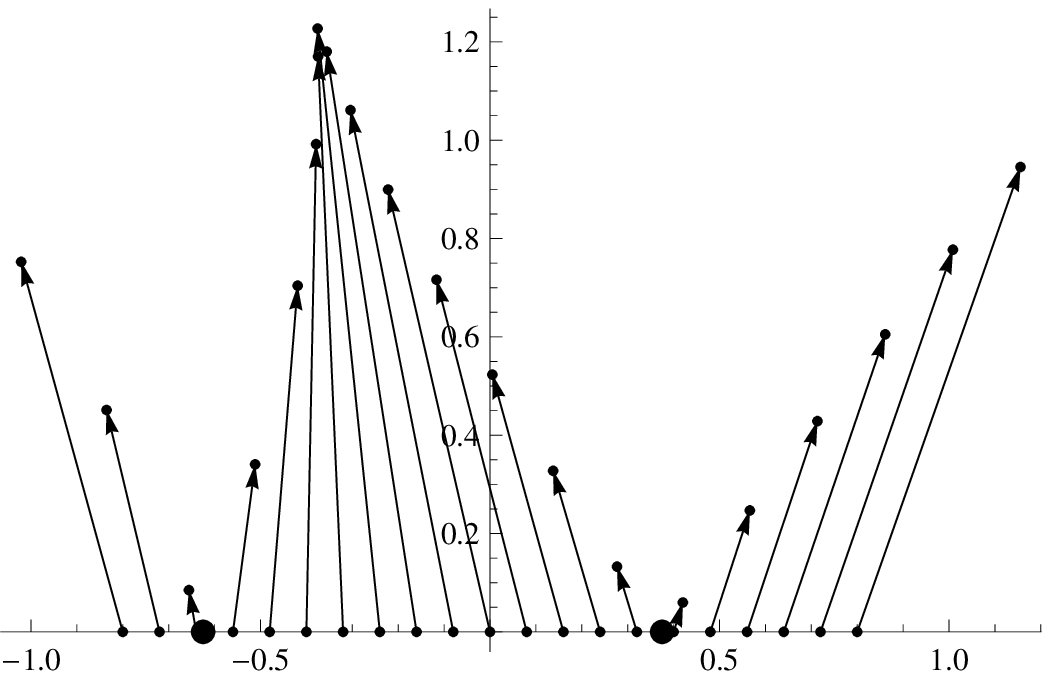}
\end{array}$
\end{center}
\caption{\footnotesize
Coordinates of the
 supersymmetric vacuum $(\Re u_{+}, \Im u_{+})$ as a function 
of the metastable vacuum $u_0$ for $N_f=2$,
with $m=0.05 \Lambda$ (left) and $m=m_c=0.5 \Lambda$ (right).
} \label{nf2e}
\end{figure}

The tunneling
rate can be estimated by Eq. (\ref{tunnel}).
The decays to all the supersymmetric vacua 
$u_{susy}=u_{s1},u_{s2},u_{s3},u_{\pm}$
must be checked to achieve a long-lived vacuum.
In figure \ref{nf2e}  $u_+$ is shown 
as a function of $u_0$;  $u_+$
and $u_0$ are almost coincident just for
$u_0 \rightarrow u_{s1},u_{s2},u_{s3}$.
This is the only limit in which it is problematic to achieve 
a long-lived metastable vacua.

It is possible to generate a local minimum nearby
the Argyres-Douglas point, 
but the allowed $(\delta \alpha, \delta \beta)$
is rather small nearby this point.
In the numerical example that we considered, we obtain that $\delta \alpha, \delta \beta \propto (u_0-u_{AD})^3$.
Nearby a non-conformal supersymmetric vacuum instead we obtain that
$\delta \alpha, \delta \beta \propto (u_0-u_{\rm susy})$.
Of course in these limits 
the parameter $\mu$ must be  very small
 in order to assure a long life to the metastable vacua.

\section{Comments on direct gauge mediation}

Direct gauge mediation 
is a well studied topic in the framework
of models which exhibit spontaneous symmetry breaking
(see for example \cite{ogm-review,KOOreview} for reviews).
Many models of direct gauge mediation
 based on generalizations of the O'Raifeartaigh model
have anomalously light gauginos in comparison to the sfermions,
 even in the absence of an R-symmetry;
 as discussed in \cite{ks}, the underlying reason for which
 gaugino masses vanish at the leading-order in SUSY-breaking is
 due to the fact that no unstable region exists in the pseudo-moduli space.
Phenomenologically gauginos whose mass scale is
lighter than the electroweak scale are very likely ruled out;
thus sfermions need to be made rather heavy;
 this has his own aesthetic problems
  because heavy sfermions induce a large correction to the Higgs mass
and reintroduce the hierarchy issue.
This is a feature also of direct gauge mediation from the ISS model \cite{ISS},
because in this case the metastable vacuum is absolutely stable in the effective
low energy description (supersymmetry is restored just due to non-perturbative effects).
A possible way to avoid light gauginos is to
consider uplifted vacua \cite{uplifted}, which are vacua
of even higher energy compared to the lowest supersymmetry breaking vacuum;
in their presence an unstable region in pseudo-moduli space becomes allowed.
It is interesting that
in the class of  perturbed $\mathcal{N}=2$ theories discussed in this paper
 it is also possible to obtain gaugino masses at the leading order in 
 SUSY-breaking; this is not in contradiction with \cite{ks}, because the metastable 
vacua that we consider are not absolutely stable in any low energy approximation 
(see also \cite{syy} for a discussion).

In the weakly coupled region the fields $Q,\tilde{Q}$
can be identified with the messengers.
For $N_c=2$ the so called
spurion of supersymmetry breaking corresponds to the adjoint
field $\Phi=a(u_0) \, \sigma_3$, where $\sigma_3$ is a Pauli matrix.
If the squark masses are set to zero,  ordinary gauge mediation (OGM) 
is realized. If the squark masses are
not zero, the gauge mediation mechanism
 is in the more general class studied in \cite{eogm};
 the gaugino and the sfermion masses are:
\beq
m_\lambda= \frac{\alpha_r}{4 \pi} \Lambda_G \, , \qquad
m_{\tilde{f}}^2 = 2 C_{\tilde{f}} \left( \frac{\alpha_r}{4 \pi} \right)^2 \Lambda_S^2 \, ,
\label{eogmeq}
\eeq
where
\beq \Lambda_G = F^a \left( \partial_a (\log \det \mathcal{M}) \right) \, , \qquad
 \Lambda_S^2 = \frac{1}{2} |F^a|^2 \frac{\partial^2}{\partial a \partial \bar{a}}
\sum_i \left( \log |\mathcal{M}_i|^2 \right)^2 \, ,  \label{eogmeqb}
 \eeq
and $\alpha_r=g_F^2/(4 \pi)$, where $g_F$ is the gauge coupling
at the messenger scale and $C_{\tilde{f}}$ is the quadratic Casimir
of the representation of the sfermion $\tilde{f}$. 

Consider for example the case with $N_c=N_f=2$
and with two identical masses for the hypermultiplets $m_1=m_2=m$.
For $m=0$ the theory has an enhanced $SO(4)$ flavor global symmetry;
for $m \neq 0$ this symmetry is broken to $SU(2)_F \times U(1)_F$,
where the $U(1)_F$ corresponds to a squark  number. 
The BPS mass formula for a state with electric and magnetic charge
$(n_e,n_m)$ and with $U(1)_F$ charge $s$ is:
\beq M_{BPS}= |\sqrt{2} n_m a_D - \sqrt{2} n_e a + s \, m |\, .\eeq
The $U(1)_F$  symmetry is  gauged and coupled to 
a an external sector, with coupling constant $\alpha_r$. 
The matrix $\mathcal{M}=\sqrt{2} \, a(u_0) \, \sigma_3 + m$ is the messenger mass matrix.
A direct evaluation gives:
\beq
\Lambda_G=F^a \frac{4a}{2 a^2-m^2} \, , \qquad
\Lambda_S^2=|F^a|^2 \frac{4 (2 |a|^2 +|m|^2)}{|2 a^2-m^2|^2} \, .  \label{eogmeqQ}
\eeq
In the $m \rightarrow 0$ limit, OGM is recovered;
the effective number of messengers $N_{eff}$ is 
\beq  N_{eff}=  \frac{\Lambda_G^2}{\Lambda_S^2}  =2 \, \label{accio}.\eeq
This shows that the gauginos are not anomalously
light in comparison to the sfermions for the metastable vacua
in the weakly coupled regime.

In the strongly coupled region of the moduli space
the fields $Q,\tilde{Q}$ can not be identified any more with the messengers;
in this regime monopoles and dyons which carry flavor quantum numbers
become lighter than the squarks, which also can become unstable particles due to
the crossing of a curve of marginal stability.
 The calculation of the gauge mediation masses in principle
 requires the calculation of the current-current correlators
 of the global symmetries, in the formalism introduced in \cite{mss}.
 An explicit expression 
 for the gaugino masses at the leading order in SUSY-breaking
  in perturbed $\mathcal{N}=2$ theories was found in  \cite{ooguri-mediation},
  for generic $N_f$ and $N_c$.
  In this more general case the Coulomb branch can parameterized
  by $N_c$ eigenvalues $a_k$, with the constraint $\sum a_k =0$. 
In order to facilitate the computation of the gaugino masses, in  \cite{ooguri-mediation}
the global symmetry  $U(1) _F \times SU(N_f)$ is gauged by introducing
  a full $\mathcal{N}=2$ vector hypermultiplet;
in this way the mass parameters $m_a$ of the original $SU(N_c)$ theory
get identified with the adjoint scalar eigenvalues
 of the $\mathcal{N}=2$ $U(1)_F \times SU(N_f)$ vector hypermultiplet.
 In the limit in which the spectator 
 $U(1)_F \times SU(N_f)$
 gauge couplings are small,
 the gaugino mass matrix  is
\beq
 (m_\lambda^{GGM})_{ab}=
g_F^2 \frac{i}{8 \pi} (F^i \mathcal{F}_{iab} - 
(\mathcal{F}_{ijm} F^m)^{-1} \mathcal{F}_{aik} 
 \mathcal{F}_{bjl} F^k F^l   )  \, , \label{ggmeq}
\eeq 
where $\mathcal{F}(a_k,m_a)$ is the prepotential
and $F^k$ is the F-term of the field $a_k$.  
Subscripts under $\mathcal{F}$ denote differentiations;
 the indices $i,j,k,l,m$ correspond to the eigenvalues $\Phi^i$
 of the adjoint field, while the indices $a,b$ correspond to
 the mass matrix eigenvalues $m_a$.

\begin{figure}[h]
\begin{center}
\leavevmode
\epsfxsize 8 cm
\epsffile{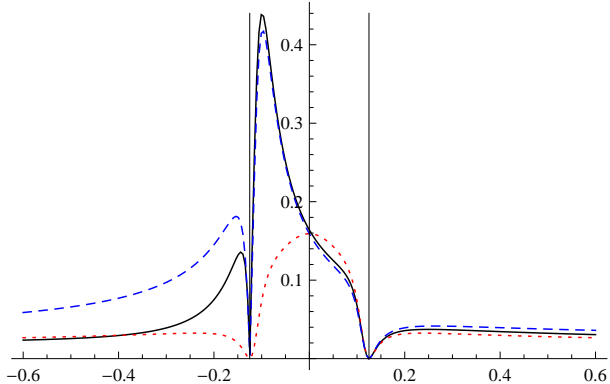}
\end{center}
\caption{\footnotesize  
The gaugino mass $m_\lambda^{\rm GGM}$, as computed from the general gauge mediation 
expression Eq.~(\ref{massina}), is shown in the solid line in units of $ g_F^2 \mu$
as a function of $u_0$. 
The mass $m_\lambda^{\rm W}$
 as computed from Eq.~(\ref{appr1}) is shown in the dotted line;
 this is a good approximation in the
 semiclassical region of the moduli space, for $u_0 \rightarrow \pm \infty$.
 The mass $m_\lambda^{\rm D}$
 as computed from Eq.~(\ref{appr2}) is shown in the dashed line; 
 this gives a good approximation inside the marginal stability curve  ($u_D<u_0<u_M$).
 The location of $u_D$ and $u_M$ is shown by the vertical lines.}
\label{gauginofig}
\end{figure} 
 
We can then apply Eq. (\ref{ggmeq})
to the gauge mediation
 of the $U(1)_F$ symmetry of the $N_f=N_c=2$ theory;
 the mass of the  gaugino then is
 \beq
 m_\lambda^{\rm GGM}= g_F^2 \left| F^a \mathcal{A} \right| \, ,
 \qquad
 \mathcal{A}= \frac{i}{8 \pi} 
 \left( \frac{\partial^2 a_D }{\partial m^2} 
-\left(\frac{\partial \tau_e}{\partial m}\right)^2 \frac{\partial a}{\partial \tau_e} \, \right)  \, , \label{massina}
\eeq 
where $F^a$ is the F-term for the field $a$:
\beq
F^a=\frac{1}{\im \, \tau_e} \frac{d \bar{W}}{d \bar{u}} \frac{d \bar{u}}{d \bar{a}} \, .
\eeq

In the following we will restrict to the case $m=0$.
In this case the structure of the singularities 
is identical to the $N_f=0$ case \cite{sw2}. There is a dyon singularity
at $u_D=-\Lambda^2/8$;
in correspondence of this vacuum two dyon states (which are $SU(2)_F$ singlets)
with electric and magnetic charges  $(n_e,n_m)=(1,-1)$ and 
with $U(1)_F$ charge $\pm 1$ become massless.
There is a monopole singularity at $u_M=\Lambda^2/8$;
for this value an $SU(2)_F$  doublet of monopoles
with $n_m=1$ and with zero $U(1)_F$ charge becomes massless.

It is interesting to compare the exact expression
in Eq.~(\ref{massina}) with the
semiclassical formula Eq.~(\ref{eogmeqQ}), which
takes into account just the contribution of the $Q,\tilde{Q}$ messengers:
  \beq m_\lambda^{\rm W}= \frac{g_F^2}{8 \pi^2}
   \left| \frac{F^a}{a}  \right| \, .  \label{appr1} \eeq
  
  In the neighborhood of the moduli space singularity
 at $u_D=-\Lambda^2/8$, another approximation can be used;
  nearby this singularity, two dyons (with $(n_e,n_m)=(1,-1)$ and with
  global $U(1)_F$ charge $\pm 1$) become  almost massless. In this limit
   these dyons give the dominant contribution to 
  the gauge mediation masses; the expressions (\ref{eogmeq} ,\ref{eogmeqb}) can be
  used, by integrating in the messengers in the form of dyon superfields 
  $D,\tilde{D}$\,\footnote{ A canonical K\"ahler potential
   is used for the dyonic fields $D, \tilde{D}$. 
  This is justified nearby the moduli space singularity 
  $u_D$, because in this region the dual gauge coupling is weak.
  In general, there could be terms in the K\"ahler potential 
  which mix $D,\tilde{D}$ with $u$.
  These terms (whose form is much restricted by $\mathcal{N}=2$ supersymmetry)
  could in principle induce a non-zero messenger supertrace 
  and make the calculation of the soft masses sensitive to physics at scales higher than
  the messenger scale  \cite{potr}.}, which  couple to the adjoint field  with the
  superpotential $\tilde{W}=\sqrt{2}  (a+a_D)  D \tilde{D}$.
  The expression for the gaugino mass in this limit then is:
    \beq m_\lambda^{\rm D}= \frac{g_F^2}{ 8 \pi^2}
    \left| F^a \frac{1+\tau_e}{a+a_D}  \right| \, .  \label{appr2}\eeq

The result of a numerical calculation for the theory with $m=0$
is shown in figure \ref{gauginofig}.
For each point of the moduli space, the coefficients $(\alpha_0,\beta_0)$
are computed; these coefficients specify
the superpotential $W$ used for each point in the moduli space.
Then  the gaugino mass is calculated using Eq.~(\ref{massina}),
see the solid line in figure \ref{gauginofig}.
In the weakly coupled region of the moduli space $u_0 \rightarrow \pm \infty$
the approximation Eq. (\ref{appr1}) can be used
  (see the dotted line in figure \ref{gauginofig}).
 The approximation in Eq. (\ref{appr2}) is plotted
 in the dashed curve; this gives a good approximation
 in the region $u_D<u_0<u_M$.

\begin{figure}[h]
\begin{center}
\leavevmode
\epsfxsize 8 cm
\epsffile{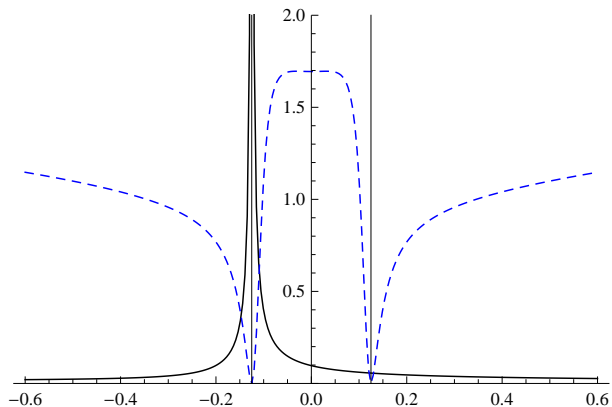}
\end{center}
\caption{\footnotesize
In the solid line, the absolute value of the coefficient
 $\mathcal{A}$ defined in Eq. (\ref{massina}) is shown  as a function of $u_0$ in units of $1/\Lambda$;
it diverges for $u_D=-\Lambda^2/8$, where a dyon charged under $U(1)_F$ becomes massless.
In the dashed line, the absolute value of the
F-term $F^a$ is shown in units of $\mu \Lambda$;
it is zero in correspondence of the supersymmetric vacua $u_{M,D}=\pm \Lambda^2/8$.
From this figure we can check that in correspondence of $u_D$ there 
is a massless particle charged under $U(1)_F$.
}
\label{gauginofig2}
\end{figure} 

The gaugino mass is proportional to the product
between the F-term and the expression $\mathcal{A}$,
as defined in defined in Eq. (\ref{massina}).
The F-term $F^a$ tends to zero as $u$ approaches the value  of the
supersymmetric vacua $u_{M,D}=\pm \Lambda^2/8$ (see figure \ref{gauginofig2}).
The coefficient $\mathcal{A}$ tends to infinity for $u_D=-\Lambda^2/8$;
this is due to the fact that in  this vacuum
some particles charged under the global $U(1)_F$ symmetry 
become massless.

 As shown in figure  \ref{gauginofig}, Eq. (\ref{appr2})
     gives a rather good approximation  for  the gaugino masses
    in all the strong coupling region with $u_D<u_0<u_M$.
   The reason for which this formula works so well in this
   region of the moduli space is probably due to the fact that we are inside
   of the marginal stability curve; in this region the only stable 
   BPS states are the monopole (which is uncharged under $U(1)_F$) 
   and the $(1,-1)$ dyons \cite{bfmarginalstab}. The relevant gauge mediation
   physics is captured by the contribution of the dyons, evaluated
   as at weak coupling.
   It is then natural to use this approximation also for the sfermions
   masses, which are proportional to
   \beq 
   \Lambda_S^2 = 2 |F^a|^2 \frac{|1+\tau_e|^2}{|a+a_D|^2} =  \frac{\Lambda_G^2}{2} \, .
   \eeq
 This is the same result as in the weakly coupled regime in Eq (\ref{accio});  
this suggests that the ordinary gauge mediation
relation $N_{eff}=2$ is satisfied with good approximation
also in the strong coupling region $u_D<u_0<u_M$.    
 The gaugino and the sfermions masses are then of comparable order
 also in this regime.

\section{Conclusions}

In this note we studied the 
issue of metastable vacua in $\mathcal{N}=2$ theories
perturbed by the superpotential in Eq.~(\ref{tumiperturbi}).
The allowed region of the parameters $(\alpha,\beta)$
in order to obtain a metastable vacuum on the Coulomb branch
was determined in some examples with $N_c=2$ and different matter content.
A general feature
in the asymptotically free cases
 is that the parameters must be considerably fine-tuned
for large $u_0>>\Lambda$, while in the strongly coupled region a smaller
degree of fine tuning is needed. Another feature is that it is more difficult
to generate metastable vacua in the conformal setting;
in order to achieve this in the $N_f=4$  and in the $\mathcal{N}=4$ theories
the conformal symmetry must be explicitly broken also
by a mass term for some of the hypermultiplets.
Also, we find that
it is hard to to achieve a metastable vacuum  nearby an
infrared Argyres-Douglas conformal point; in the explicit example
that we studied we found that $\delta \alpha, \delta \beta$
vanish as $(u_0-u_{AD})^3$, which is  stronger than
nearby the other supersymmetric vacua, where we find that
 $\delta \alpha, \delta \beta$ vanish in a linear way in $(u_0-u_{susy})$.

Direct gauge mediation can be implemented
with sizable gaugino masses already at the leading order in
SUSY breaking.  In particular, in the case of zero hypermultiplets
masses, ordinary gauge mediation is realized.

\section*{Acknowledgments}

We are grateful to Amit Giveon, Zohar Komargodski
and Stefan Theisen for useful discussions.
The work of E. Rabinovici was partially supported by the Humbodlt foundation, a DIP grant H, 52, the Einstein Center at the Hebrew University, the American-Israeli Bi-National Science Foundation and the Israel Science Foundation Center of Excellence.
The work of  R. Auzzi was partially supported by a DIP grant H, 52, the Einstein Center at the Hebrew University, the American-Israeli Bi-National Science Foundation and the Israel Science Foundation Center of Excellence.

\section*{Appendix. The weakly coupled limit } 

\label{weakcoupling}

For $N_f=0 \ldots 3$, 
in the weakly coupled region of the moduli space, 
$u >> \Lambda^2, m_i^2$,
the form of $(a_D,a)$ is determined by the one-loop $\beta$ function,
which (setting $\Lambda=1$) leads to
\beq a(u) \approx \sqrt{ \frac{u}{2}} \, , \qquad
a_D(u) \approx  \frac{i}{4 \pi}  (4-N_f)  \sqrt{2 u } \log u \, .\eeq
The moduli space metric is
\beq g=\frac{4-N_f}{16 \pi} \, \frac{2+\log |u|}{|u|} \, . \eeq
We can then write a compact expression the four functions
 $(\alpha_0,\beta_0,\delta \alpha, \delta \beta)$:
\beq \alpha_0= -\frac{8 + 5 \log |u_0|^2}{u_0 (44+15 \log |u_0|^2)}\approx -\frac{1}{3 u_0}
 \, ,  \eeq
\[ \beta_0=\frac{4+3 \log |u_0|^2}{3 u_0^2 (44+15 \log |u_0|^2)}
\approx \frac{1}{15 u_0^2} \, , \]
\[ \delta \alpha= - \frac{8 (18 + 5 \log |u_0|^2)}
{\bar{u}_0 (4+ \log |u_0|^2) (44 + 15 \log |u_0|^2)^2}  \approx
- \frac{8}{45 \bar{u}_0 (\log |u_0|^2)^2} \, ,
\]
\[ 
\delta \beta = \frac{16(10+ 3 \log |u_0|^2)}
{ 3 |u_0|^2 (4+ \log |u_0|^2) (44 + 15 \log |u_0|^2)^2} \approx
\frac{16}{225 |u_0|^2  (\log |u_0|^2)^2}
\, .
\]
A general feature of the weakly-coupled region is that
the perturbation parameters $(\alpha,\beta)$ must
be small and rather fine tuned 
(with  $\beta \approx  3/5 \alpha^2$)
in order to achieve metastability.


\begin{thebibliography}{100}

\bibitem{ISS}
  K.~A.~Intriligator, N.~Seiberg and D.~Shih,
  JHEP {\bf 0604} (2006) 021
  [arXiv:hep-th/0602239].

\bibitem{ISreview}
  K.~A.~Intriligator and N.~Seiberg,
  Class.\ Quant.\ Grav.\  {\bf 24} (2007) S741
  [arXiv:hep-ph/0702069].
 
\bibitem{KOOreview}
  R.~Kitano, H.~Ooguri and Y.~Ookouchi,
  arXiv:1001.4535 [hep-th].

\bibitem{Sdual}
  N.~Seiberg,
  Nucl.\ Phys.\  B {\bf 435}, 129 (1995)
  [arXiv:hep-th/9411149].

\bibitem{ITYY}
 K.~I.~Izawa, F.~Takahashi, T.~T.~Yanagida and K.~Yonekura,
  Phys.\ Rev.\  D {\bf 80}, 085017 (2009)
  [arXiv:0905.1764 [hep-th]];
T.~T.~Yanagida and K.~Yonekura,
  arXiv:1002.4093 [hep-th].


\bibitem{AGMS}
 A.~Amariti, L.~Girardello, A.~Mariotti and M.~Siani,
  arXiv:1003.0523 [hep-th].


\bibitem{sw1}
  N.~Seiberg and E.~Witten,
  Nucl.\ Phys.\  B {\bf 426} (1994) 19
  [Erratum-ibid.\  B {\bf 430} (1994) 485]
  [arXiv:hep-th/9407087].

\bibitem{sw2}
  N.~Seiberg and E.~Witten,
  Nucl.\ Phys.\  B {\bf 431} (1994) 484
  [arXiv:hep-th/9408099].


\bibitem{OOP}
  H.~Ooguri, Y.~Ookouchi and C.~S.~Park,
  Adv.\ Theor.\ Math.\ Phys.\  {\bf 12} (2008) 405
  [arXiv:0704.3613 [hep-th]].

\bibitem{MOOP}
  J.~Marsano, H.~Ooguri, Y.~Ookouchi and C.~S.~Park,
  Nucl.\ Phys.\  B {\bf 798} (2008) 17
  [arXiv:0712.3305 [hep-th]].

\bibitem{pastras1}
  G.~Pastras,
  arXiv:0705.0505 [hep-th].

\bibitem{pastras2}
  E.~Katifori and G.~Pastras,
  arXiv:0811.3393 [hep-th].

  
 \bibitem{amos}
  M.~Arai, C.~Montonen, N.~Okada and S.~Sasaki,
  Phys.\ Rev.\  D {\bf 76} (2007) 125009
  [arXiv:0708.0668 [hep-th]].
    M.~Arai, C.~Montonen, N.~Okada and S.~Sasaki,
  JHEP {\bf 0803} (2008) 004
  [arXiv:0712.4252 [hep-th]];

 \bibitem{others} 
  L.~Mazzucato, Y.~Oz and S.~Yankielowicz,
  JHEP {\bf 0711}, 094 (2007)
  [arXiv:0709.2491 [hep-th]];
 J.~Marsano, K.~Papadodimas and M.~Shigemori,
  Nucl.\ Phys.\  B {\bf 804}, 19 (2008)
  [arXiv:0801.2154 [hep-th]];
  L.~Hollands, J.~Marsano, K.~Papadodimas and M.~Shigemori,
  JHEP {\bf 0810}, 102 (2008)
  [arXiv:0804.4006 [hep-th]].

\bibitem{scales}
 E.~Rabinovici, B.~Saering and W.~A.~Bardeen,
 Phys.\ Rev.\  D {\bf 36}, 562 (1987);
 D.~J.~Amit and E.~Rabinovici,
 Nucl.\ Phys.\  B {\bf 257}, 371 (1985);
 D.~S.~Berman and E.~Rabinovici,
 arXiv:hep-th/0210044.

 \bibitem{adapsw}
  P.~C.~Argyres and M.~R.~Douglas,
  Nucl.\ Phys.\  B {\bf 448} (1995) 93
  [arXiv:hep-th/9505062];
  P.~C.~Argyres, M.~Ronen Plesser, N.~Seiberg and E.~Witten,
  Nucl.\ Phys.\  B {\bf 461} (1996) 71
  [arXiv:hep-th/9511154].

\bibitem{ogm-review}
  G.~F.~Giudice and R.~Rattazzi,
  Phys.\ Rept.\  {\bf 322}, 419 (1999)
  [arXiv:hep-ph/9801271].


\bibitem{bilalrev}
  A.~Bilal,
  arXiv:hep-th/9601007.

\bibitem{triang}
  M.~J.~Duncan and L.~G.~Jensen,
  Phys.\ Lett.\  B {\bf 291}, 109 (1992).

\bibitem{egr}
M.~B.~Einhorn, G.~Goldberg and E.~Rabinovici,
  Nucl.\ Phys.\  B {\bf 256}, 499 (1985).


\bibitem{ferrari}
  F.~Ferrari,
  Nucl.\ Phys.\  B {\bf 501} (1997) 53
  [arXiv:hep-th/9702166].

  \bibitem{dkm1}
  N.~Dorey, V.~V.~Khoze and M.~P.~Mattis,
  Nucl.\ Phys.\  B {\bf 492} (1997) 607
  [arXiv:hep-th/9611016].

\bibitem{dkm2}
  N.~Dorey, V.~V.~Khoze and M.~P.~Mattis,
  Phys.\ Lett.\  B {\bf 396} (1997) 141
  [arXiv:hep-th/9612231].
  
  
  
\bibitem{bf2}
  A.~Bilal and F.~Ferrari,
  Nucl.\ Phys.\  B {\bf 516} (1998) 175
  [arXiv:hep-th/9706145].


\bibitem{ks}
  Z.~Komargodski and D.~Shih,
  JHEP {\bf 0904} (2009) 093
  [arXiv:0902.0030 [hep-th]]. 
  

  
  \bibitem{uplifted}
  R.~Kitano, H.~Ooguri and Y.~Ookouchi,
  Phys.\ Rev.\  D {\bf 75} (2007) 045022
  [arXiv:hep-ph/0612139];
  A.~Giveon, A.~Katz and Z.~Komargodski,
  JHEP {\bf 0907} (2009) 099
  [arXiv:0905.3387 [hep-th]];
  S.~A.~Abel, J.~Jaeckel and V.~V.~Khoze,
  Phys.\ Lett.\  B {\bf 682} (2010) 441
  [arXiv:0907.0658 [hep-ph]].

\bibitem{syy}  
  S.~Shirai, M.~Yamazaki and K.~Yonekura,
  arXiv:1003.3155 [hep-ph].

\bibitem{eogm}
  C.~Cheung, A.~L.~Fitzpatrick and D.~Shih,
  JHEP {\bf 0807}, 054 (2008)
  [arXiv:0710.3585 [hep-ph]].

\bibitem{mss}
  P.~Meade, N.~Seiberg and D.~Shih,
  Prog.\ Theor.\ Phys.\ Suppl.\  {\bf 177}, 143 (2009)
  [arXiv:0801.3278 [hep-ph]].


\bibitem{ooguri-mediation}
  H.~Ooguri, Y.~Ookouchi, C.~S.~Park and J.~Song,
  Nucl.\ Phys.\  B {\bf 808} (2009) 121
  [arXiv:0806.4733 [hep-th]].
 
 \bibitem{potr}
  E.~Poppitz and S.~P.~Trivedi,
  Phys.\ Lett.\  B {\bf 401}, 38 (1997)
  [arXiv:hep-ph/9703246].
  
\bibitem{bfmarginalstab}  
  F.~Ferrari and A.~Bilal,
  Nucl.\ Phys.\  B {\bf 469}, 387 (1996)
  [arXiv:hep-th/9602082];
   A.~Bilal and F.~Ferrari,
  Nucl.\ Phys.\  B {\bf 480}, 589 (1996)
  [arXiv:hep-th/9605101].
  
  
\end{thebibliography}
\end{document}